\documentclass{aa}
\usepackage{graphicx}
\usepackage{xcolor}
\usepackage{multicol}
\usepackage{natbib}
\usepackage{subcaption}
\usepackage{tabularx,ragged2e,booktabs}
\usepackage[]{algorithm2e,algpseudocode}
\usepackage{amsmath}
\usepackage{amssymb,amsfonts,textcomp}
\usepackage{txfonts}
\usepackage{subcaption,siunitx,booktabs}
\usepackage[titletoc,toc,title]{appendix}
 
\newlength\myindent
\newcommand{\resp}{resp{.}}
\newcommand{\ie}{i{.}e{.}~}
\newcommand{\eg}{e{.}g{.}~}
\newcommand{\eq}{Eq{.}~}                 
\newcommand{\eqs}{Eqs{.}~}
\newcommand{\fg}{Fig{.}~}
\newcommand{\fgs}{Figs{.}~}
\newcommand{\sct}{Sect{.}~}

\usepackage{tikz}
\usetikzlibrary{shapes,arrows,fit,calc,positioning}
 
\begin{document}

   \title{Inferring the photometric and size evolution of galaxies from image simulations}
 
   \subtitle{I. Method}
 
   \author{S\'ebastien Carassou\inst{1},
          Val\'erie de Lapparent\inst{1},
          Emmanuel Bertin\inst{1},
          Damien Le Borgne\inst{1}
          }
 
   \authorrunning{Carassou, de Lapparent, Bertin \& Le Borgne}
   \titlerunning{Evolution of galaxies from image simulations}
   \institute{$^1$ Institut d'Astrophysique de Paris, 98 bis bd Arago, 75014 Paris, France, CNRS, UMR 7095 et Sorbonne Universit\'es, UPMC Univ Paris 6  \\
              \email{lapparent@iap.fr}
         }
 
   \date{Received ; accepted }
 
 
  \abstract
   {Current constraints on models of galaxy evolution rely on morphometric catalogs extracted from multi-band photometric surveys. However, these catalogs are altered by selection effects that are difficult to model, that correlate in non trivial ways, and that can lead to contradictory predictions if not taken into account carefully.}
   {To address this issue, we have developed a new approach combining parametric Bayesian indirect likelihood (pBIL) techniques and empirical modeling with realistic image simulations that reproduce a large fraction of these selection effects. This allows us to perform a direct comparison between observed and simulated images and to infer robust constraints on model parameters. }
   {We use a semi-empirical forward model to generate a distribution of mock galaxies from a set of physical parameters. These galaxies are passed through an image simulator reproducing the instrumental characteristics of any survey and are then extracted in the same way as the observed data. The discrepancy between the simulated and observed data is quantified, and minimized with a custom sampling process based on adaptive Monte Carlo Markov Chain methods.}
   {Using synthetic data matching most of the properties of a Canada-France-Hawaii Telescope Legacy Survey Deep field, we demonstrate the robustness and internal consistency of our approach by inferring the parameters governing the size and luminosity functions and their evolutions for different realistic populations of galaxies. We also compare the results of our approach with those obtained from the classical spectral energy distribution fitting and photometric redshift approach.}
   {Our pipeline infers efficiently the luminosity and size distribution and evolution parameters with a very limited number of observables (three photometric bands). When compared to SED fitting based on the same set of observables, our method yields results that are more accurate and free from systematic biases.}
   \keywords{galaxies: evolution -- galaxies: bulges -- galaxies: spiral -- galaxies: luminosity function, mass function -- galaxies: statistics -- methods: numerical}
 \maketitle
%
\section{Introduction} 
 
 
  During the last decades our understanding of galaxy formation and evolution has been largely shaped by the results of deep multicolor photometric surveys. We can now extract the spectro-photometric properties of millions of galaxies, over large volumes that cover more than ten billion years of cosmic history. Despite this wealth of data, we are still incapable of deriving strong constraints on the free parameters of current semi-analytic models that describe quantitatively how galaxies evolve in color, size, and shape from their high redshifts counterparts. The main reason is that, missing physical ingredients in our models aside, the galaxy catalogs derived from surveys are often incomplete. 

First of all, surveys are limited in flux. Consequently, intrinsically faint sources tend to be under-represented because they are above the limiting magnitude only at small distances. This effect, called Malmquist bias  \citep{Malmquist1920},
 introduces correlations between probably non-correlated variables, mainly distance and other parameters such as luminosity (e.g., \citealt{Singal2014}). Additionally, some galaxies overlap and may be blended into single objects. Source confusion \citep{Condon1974}, caused by unresolved faint sources blended by the point spread function, can act as a signal at the detection limit and also affects number counts in a non-trivial way. Moreover, source confusion affects background estimation by adding a non-uniform component to the background noise, which is correlated with the spatial distribution of unresolved sources \citep{helou90}. Statistical fluctuations in flux measurements give rise to the Eddington bias \citep{Eddington1913}. As galaxy number counts increase as a power of the flux, there are more overestimated fluxes for faint sources than underestimated fluxes for bright sources. This results in a general increase in the number of sources detected at a given flux \citep{HASINGER2000,Loaring2005}. Because of the cosmological dimming, the bolometric surface brightness of galaxies gets dimmer with increasing redshift proportionally to $(1+z)^{-4}$ \citep{Tolman1934}, which makes many faint extended sources undetectable. Finally, stellar contamination affects the bright end of the source counts (e.g., \citealt{Pearson2014}). 
 
 Apparent magnitudes in catalogs also have to be corrected for Galactic extinction (e.g., \citealt{Schlegel1997}), and to account for redshift effects, K-corrections (\citealt{Hogg2002}) that are sensitive to galaxy spectral type must be applied on the magnitudes of high-redshift galaxies (e.g., \citealt{Ramos2011}). Both corrections, however, are applied only after the sample is truncated at its flux limit, which causes biases at the survey limit.  Inclination-dependent internal absorption from dust lanes in the disk of galaxies also tends to draw a fraction of edge-on spirals below the survey flux limit (e.g., \citealt{Kautsch2006}). Because of these various selection effects, that correlate in ways that are poorly understood, and that may be spatially variable over the field of view of the survey, observations undergo complex selection functions that are difficult to treat analytically, and the resulting catalogs tend to be biased towards intrinsically brighter, compact, and low dust content sources.

The determination of the luminosity function (LF) of galaxies, a fundamental tool for characterizing galaxy populations that is often used for constraining models of galactic evolution, is particularly sensitive to these biases. 
As input data, analyses use catalogs containing the photometric properties, such as apparent magnitudes, of a selected galaxy sample. LF estimation requires the knowledge of the absolute magnitude of the sources, which itself depends upon the determination of their redshift. The number density per luminosity bin can be determined by a variety of methods, parametric or non-parametric, described in detail in \citet{Binggeli1988}, \citet{Willmer1997}, and \citet{Sheth2007}. The resulting distribution is usually fitted by a Schechter function \citep{Schechter1976}, but other functions are sometimes required (e.g., \citealt{driver96}, \citealt{blanton05}). The Schechter function is characterized by three parameters:  $\phi^*$ the normalization density, $\alpha$ the faint end slope, and $M^*$ a characteristic absolute magnitude. The LF at z $\sim$ 0 is presently well constrained thanks to the analysis of high-resolution spectroscopic surveys, such as the 2dF Galaxy Redshift Survey (2dFGRS, \citealt{Norberg2002}) or the Sloan Digital Sky Survey (SDSS, \citealt{Blanton2002}). There is also clear evidence that the global LF evolves with redshift, and that the LFs for different populations of galaxies evolve differently (\citealt{Lilly1995}, \citealt{Zucca2005}). 

Measuring the LF evolution is nevertheless a challenge, as high-redshift galaxies are faint, and therefore generally unsuitable for spectroscopic redshift determination, which would require prohibitive exposure times. The current solution to this problem is to use the information contained within the fluxes of these sources in some broad-band filters, in order to estimate their redshift, known as photometric redshift. This procedure has a number of biases in its own right, because the precision of photometric redshifts relies on the templates and the training set used, assumed to be representative of the galaxy populations. These biases are described  extensively in \citet{MacDonald2010}. In turn, redshift uncertainties typically result in an increase of the estimated number of low and high luminosity galaxies (\citealt{Sheth2007}).

\paragraph{The forward-modeling approach to galaxy evolution}$~$\\
 
The traditional approach when comparing the results of models to data is sometimes referred to as backward modeling (e.g., \citealt{Marzke1998}, \citealt{Taghizadeh-Popp2015}). In this scheme, physical quantities are derived from the observed data, and are then compared with the physical quantities predicted from simulations, semi-analytical models (SAM), or semi-empirical models. A more reliable technique is the forward modeling approach: a distribution of modeled galaxies are passed through a virtual telescope with all the observing process reproduced (filters, exposure time, telescope characteristics, seeing properties, as well as the cosmological and instrumental biases described above), and a direct comparison is made between simulated and observed datasets. The power of this approach comes from the fact that theory and observation are compared in the observational space: the same systematic errors and selection effects affect the simulated and observed data. \citet{Blaizot2003} were the first to introduce realistic mock telescope images from light cones generated by SAMs. \citet{Overzier2012} extended this idea by constructing synthetic images and catalogs from the Millenium Run cosmological simulation including detailed models of  ground-based and space telescopes. More recently, \citet{Taghizadeh-Popp2015} used semi-empirical modeling to simulate Hubble Deep Field (HDF) images, from cutouts of real SDSS galaxies with modified sizes and fluxes, and compared them to observed HDF images. Here we make the case that forward modeling can be used to perform reliable inferences on the evolution of the galaxy luminosity and size functions.
 
 
\paragraph{Bayesian inference}$~$\\
 
Standard Bayesian techniques provide a framework to address any statistical inference problem. The goal of Bayesian inference is to infer the posterior probability density function (PDF) of a set of model parameters $\theta$, given some observed data $\mathcal{D}$. This probability can be derived using the Bayes' theorem: 
\begin{align}
P(\theta | \mathcal{D}) = \frac{P(\mathcal{D} | \theta) P(\theta)}{P(\mathcal{D})},
\label{eq:bayes}
\end{align}
where $P(\mathcal{D}| \theta)$ is also called the likelihood (or the likelihood function) of the data, which gives the probability of the data given the model, $P(\theta)$ is the prior, or the probability of the model with the parameters $\theta$, and $P(\mathcal{D})$ is the evidence, which acts as a normalization constant and is usually ignored in inference problems. The posterior PDF is approximated either analytically or via the use of sampling techniques, such as Markov Chain Monte Carlo (MCMC). 

However, there are multiple cases where the likelihood is intractable or unknown, for mathematical or computational reasons, which renders classical Bayesian approaches unfeasible. In our case, it is the modeling of the selection effects that is impractical to include in the likelihood. To tackle this issue, a new class of methods, called ``likelihood-free'', have been developed to infer posterior distributions without explicit computation of the likelihood.

\paragraph{Approximate Bayesian Computation}$~$\\

One of the ``likelihood-free'' techniques is called Approximate Bayesian Computation (ABC), and was introduced in the seminal article of \citet{Pritchardetal99} for population genetics. ABC is based on repeated simulations of datasets generated by a forward model, and replaces the likelihood estimation by a comparison between the observed and synthetic data. Its ability to perform inference under arbitrarily complex stochastic models, as well as its well established theoretical grounds, have lead to its growing popularity in many fields, including ecology, epidemiology, and stereology (see \citealt{Beaumont2010} for an overview).
 
The classic ABC Rejection sampling algorithm, introduced in its modern form by \citet{Pritchardetal99}, is defined in Algorithm \ref{alg:abcrejectsampling},
\begin{algorithm}
\caption{ABC Rejection sampling algorithm}
\For{t = 1 to $T$}
    {\textbf{Repeat} \\
    \qquad Generate $\theta^*$ from the prior distribution \;
    \qquad Simulate data $\mathcal{D}^*$ from parameters $\theta^*$\;
    \textbf{until} $\rho(\eta(\mathcal{D}^*),\eta(\mathcal{D})) \leq \epsilon$ \;
    set $\theta_{(t)}=\theta^*$ \;
    }
\label{alg:abcrejectsampling}
\end{algorithm}
where $\rho$ is a distance metric built between the simulated and observed datasets, usually based on some summary statistics $\eta$, which are parameters that maximize the information contained within the datasets (for example, normally distributed datasets can be characterized using the mean and standard deviation of the underlying Gaussian distribution), and $\epsilon$ is a user-defined tolerance level $> 0$. Using the ABC algorithm with a good summary statistic and a small enough tolerance ultimately leads to a fair approximation of the posterior distribution (\citealt{Sunnaker2013}). The choices of $\rho$,$\eta$ and $\epsilon$ are highly non-trivial though, and they constitute the fundamental difficulty in the application of ABC methods as they are problem-dependent (\citealt{Marin2011}). Moreover rejection sampling is notorious for its inherent inefficiency, as  sampling directly from the prior distribution results in spending computing time simulating datasets in low-probability regions. Therefore, several classes of sampling algorithms have been developed to explore the parameter space more efficiently. Three of the most popular of them are outlined below.
 
\begin{itemize}
\item In the \textit{ABC-MCMC} algorithm (\citealt{Marjoram2003}), a point in the parameter space called a particle performs a random walk (defined by a proposal distribution or transition kernel) across the parameter space, and is only moving if the simulated dataset generated by these parameters match better the observed dataset, until it converges to a stationary distribution. As in standard MCMC procedures, the efficiency of the algorithm is largely determined by the choice of the scale of the kernel.
 
\item In the \textit{ABC Sequential Monte Carlo} parallel algorithm (ABC-SMC, \citealt{Toni2009}), samples are drawn from the prior distribution until $N$ particles are accepted, that is, those with a distance to the data $< \epsilon_0$. All accepted particles are attributed a statistical weight $\omega_0$.
The weighted particles then constitute an intermediate distribution from which another set of samples is drawn and perturbed with a fixed transition kernel, until $N$ particles satisfy the acceptance criterion: $\rho < \epsilon_1$, with $\epsilon_1 < \epsilon_0$. They are then weighted with $\omega_1$ and the process is repeated with a diminished tolerance at each step. After T iterations of this process, the particles are sampled from the approximated posterior distribution. The performance of ABC-SMC scales as $N$, where $N$ is the number of particles. Different variations of ABC-SMC algorithms have been published, each with a different weighting scheme for particles.
 
\item \textit{ABC Population Monte Carlo} (ABC-PMC, \citealt{Beaumont2008}) is similar to ABC-SMC, but differs in its adaptive weighting scheme: its transition kernel is Gaussian and based on the variance of the accepted particles in the previous iteration. This scheme requires the fewest tuning parameters of the three algorithms discussed here (\citealt{Turner2012}). But ABC-PMC is also more computationally costly than ABC-SMC, as its performance scales as $N^2$ (caused by its adaptability).
 
\end{itemize}
 
The reader is referred to \citet{Csilleryetal10}, \citet{Marin2011}, \citet{Turner2012}, \citet{Sunnaker2013}, and \citet{gutman2016} for a set of historical, methodical, and theoretical reviews of this final approach, as well as a complete description of the algorithms mentioned above.

\paragraph{Parametric Bayesian indirect likelihood}$~$ \\

Another class of likelihood-free techniques is called parametric Bayesian indirect likelihood (pBIL). First proposed by \citet{Reeves2005} and \citet{Gallant2009}, pBIL transforms the intractable likelihood of complex inference problems into a tractable one using an auxiliary parametric model that describes the simulated datasets generated by the forward model. In this scheme, the resulting auxiliary likelihood function quantifies the discrepancy between the observed and simulated data. It is used in Bayes' theorem and the parameter space is explored using a user-defined sampling procedure, in an equivalent way to a classical Bayesian technique. While sharing similarities with the previous technique, pBIL is not an ABC method in the strict sense, as it does not require an appropriate choice of summary statistics and tolerance level to compare the observed and synthetic datasets. The accuracy of the inference in the pBIL scheme is determined by how well the auxiliary model describes the data (observed and simlated). The theoretical foundations of this scheme are described extensively in \citet{drovandi2015}. 
 
\paragraph{Application of likelihood-free inference to astrophysics}$~$ \\
 
The application of likelihood-free methods to astrophysics is still rare, as noted by \citet{Cameron2012} in their review. Only lately has the potential of such techniques been considered. \citet{Schafer2012} praised the use of likelihood-free inference in the context of quasar luminosity function estimation. \citet{Cameron2012} explored the morphological transformation of high-redshift galaxies and derived strong constraints on the evolution of the merger rate in the early Universe using an ABC-SMC approach. \citet{Weyant2012} also used SMC for the estimation of cosmological parameters from type Ia supernovae samples, and could still provide robust results when the data was contaminated by type IIP supernovae. \citet{Robinetal14} constrained the shape and formation period of the thick disk of the Milky Way using MCMC as their sampling scheme, based on photometric data from the SDSS and the Two Micron All Sky Survey (2MASS). Finally \citet{Hahn2016} demonstrate the feasibility of using ABC to constrain the relationship between galaxies and their dark matter halo. The recent birth of Python packages providing sampling algorithms in an ABC framework, such as astroABC (\citealt{Jennings2016}) and ELFI (\citealt{kangasraasio2016}), which implement SMC methods, and COSMOABC (\citealt{Ishida2015}) which implements the PMC algorithm, will probably facilitate the rise of likelihood-free inference techniques in the astronomical community.
 
\paragraph{Outline of the article}$~$\\
 
To the authors' knowledge, no likelihood-free inference approaches have yet included telescope image simulation in their forward modeling pipeline, because of the difficulty in implementation as well as a prohibitive computational cost. Prototypical implementations in a cosmological context have, however, been tested by \citet{Akeretetal15} on a Gaussian toy model for the calibration of image simulations. In the present article we propose a new technique that combines the forward modeling approach with sampling techniques in the pBIL framework. In that regard, we use a stochastic semi-empirical model of evolving galaxy populations coupled to an image simulator to generate realistic synthetic images. Simulated images go through the same source extraction process and data analysis pipeline as real images. The observed and synthetic data distributions are finally compared and used to infer the most probable models.
 
This article is organized as follows: Sections \ref{sec:model} to \ref{subsec:mcmc} describe in detail the forward-modeling pipeline we propose, from model parameters to data analysis and sampling algorithm. Section \ref{sec:convergence} defines our convergence diagnostics. In Section \ref{sec:application}, we demonstrate the validity, internal consistency and robustness of our approach by inferring the LF parameters and their evolution using one realization of our model as input data. We perform these tests in two situations : a configuration where the data is a mock Canada-France-Hawaii Telescope Legacy Survey (CFHTLS) Deep image containing two populations of ellipticals and lenticulars and late-type spirals, and where the parameters to infer are the evolving luminosity function parameters for each population (Section \ref{subsec:2_pop}); and a configuration where the data is a mock CFHTLS Deep image with a single population of pure bulge elliptical galaxies, and in which the inference is performed on the evolving size and luminosity (Section \ref{subsec:1popE}). In Section \ref{sec:photoz}, we compare the results of our forward modeling approach with those of the more traditional photometric redshift approach applied to the same situation. Finally, Section \ref{sec:discussion} provides suggestions to improve the speed and accuracy of this method.
 
Throughout this article, unless stated otherwise, we adopt the following cosmological parameters: $H_0$ = $100h$.km.$s^{-1}$.Mpc$^{-1}$ with $h=1$, $\Omega_m=0.3$, $\Omega_{\Lambda}=0.7$ \citep{2003ApJS..148..175S}. Magnitudes are given in the AB system.
 

\begin{figure}
 
 \tikzstyle{decision} = [diamond, draw, fill=blue!20,
    text width=4.5em, text badly centered, node distance=2.5cm, inner sep=0pt]
\tikzstyle{block} = [rectangle, draw, fill=blue!20,
  text centered, rounded corners, minimum height=4em]
\tikzstyle{line} = [draw, thick, color=black!50, -latex]
\tikzstyle{cloud} = [draw, ellipse,fill=red!20, node distance=2.5cm,
    minimum height=2em]
\tikzstyle{c} = [rectangle, draw, minimum height=15em, minimum width=10em, dashed]
 
\begin{tikzpicture}[node distance = 2cm, auto]
    \node [cloud,align=center] (params) {Physical parameters \\ {\scriptsize Sections \ref{subsec:stuff} and \ref{sec:prior} } };
    \node [block, right of=params, node distance = 5cm, text width=6em] (modsourcecat) {multi-$\lambda$ galaxy catalog: \\ {\sc Stuff} \\ {\scriptsize Section \ref{subsec:stuff}}};
    \node [block, below of=modsourcecat, node distance = 2.5cm, text width=6em, xshift = -2.5cm] (modimage) {multi-$\lambda$ simulated image: \sc SkyMaker \\ \scriptsize Section \ref{subsec:skymaker}};
 
    \node [block, right of=modimage, node distance=3.5 cm, text width=6em] (obsimage) {multi-$\lambda$ observed image};
 
    \node [block, below of=modimage, xshift=2cm,yshift=-1.5cm,text width=6em,align=center] (sextractor) {Source extraction: {\sc SExtractor} \\ {\scriptsize Section \ref{subsec:sextractor} } };
    \node [block, below of=sextractor,align=center] (catcompress) {Catalog compression \\ {\scriptsize Sections \ref{subsec:reddynrange} and \ref{subsec:whitening}}};
    \node [block, below of=catcompress,align=center] (binning) {Binning \\ {\scriptsize Section \ref{subsec:binning}}};
    \node[c,fit=(sextractor) (catcompress) (binning)] (container) {};
 
    \node [block, below of=binning,align=center] (loglikelihood){Auxiliary likelihood \\ {\scriptsize Section \ref{sec:comparison}}};
 
    \path [line,align=center] (params) -- node {} (modsourcecat);
    \path [line] (modsourcecat) -- node [xshift=-1.8cm, yshift=0.5cm,align=center] {} (modimage);
 
    \path [line] (modimage) -- (container);
    \path [line] (obsimage) -- (container);
 
    \path [line] (sextractor) -- (catcompress);
    \path [line] (catcompress) -- (binning);
 
    \path [line] ([xshift=-1cm]container.south) -- ([xshift=-0.5cm]loglikelihood.north);
    \path [line] ([xshift=1cm]container.south) -- ([xshift=0.5cm]loglikelihood.north);
 
    \path [line] (loglikelihood) -| node [ right, text width=6em,align=center] {MCMC chain to maximize likelihood \\ {\scriptsize Section \ref{subsec:mcmc} }} (params);
\end{tikzpicture}
\caption{Summary of the workflow}
\label{fig:workflow}
\end{figure}
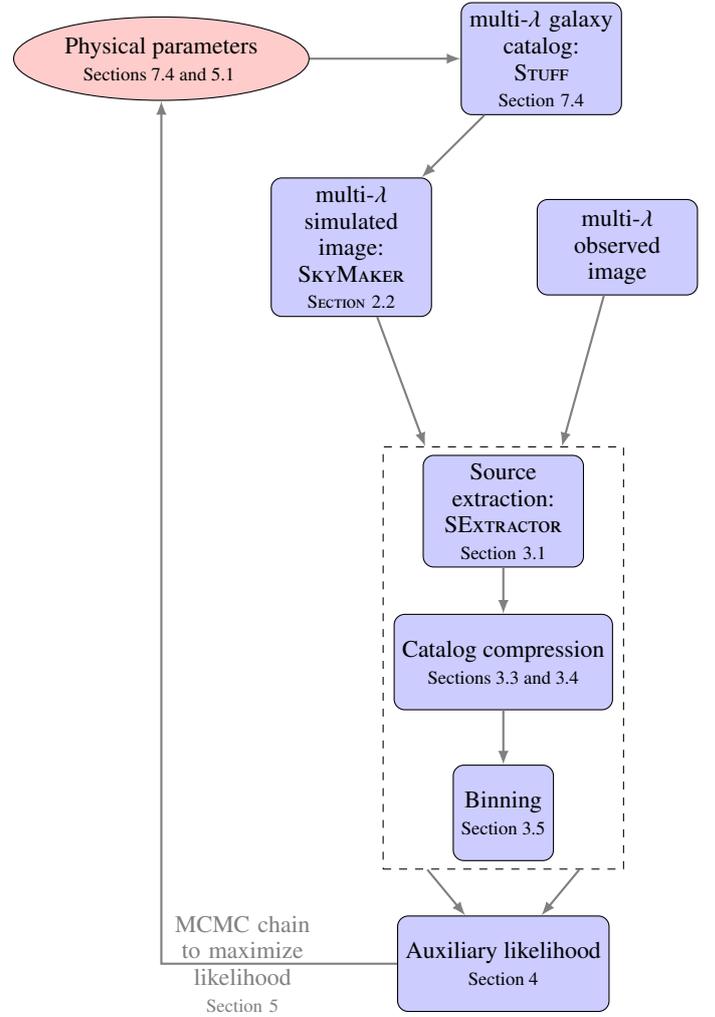

\section{Model: from parameters to image generation}
 \label{sec:model}
 
In order to infer the physical properties of galaxies from observed survey images without having to describe the complex selection effects the latter contain, we propose the following pipeline. We start from a set of physical input parameters, drawn from the prior distribution defined for each parameter. These parameters describe the luminosity and size distribution of the various populations of modeled galaxies. From this set of parameters, our forward model generates a catalog of galaxies modeled as the sum of their bulge and disk components, each with a different profile. The projected light profiles of the galaxies are determined by their inclination, the relative fraction of light contained within the bulge, and the galaxy redshift as well as the extinction of the bulge and disk components. The galaxies are randomly drawn from the luminosity function of their respective population. The catalog assumes that galaxies are randomly distributed on a virtual sky that includes the cosmological effects of an expanding universe with a cosmological constant. The survey image is simulated in every band covered by the observed survey, and reproduces all of its characteristics, such as filters transmission, exposure time, point spread function (PSF) model, and background noise model.

Then, a large number of ``simulated'' images are generated via an iterative process (a Markov chain) generating new sets of physical parameters at each iteration. Some basic flux and shape parameters are extracted in the same way from the observed and simulated images: after a pre-processing step (which is identical for observed and simulated data) where observables are decorrelated and their dynamic range reduced, the multidimensional distributions of simulated observables are directly compared to the observed distributions using a custom distance function on  binned data. 

The chain moves through the parameter space towards regions of high likelihood, that is, regions that minimize the distance between the modeled and observed datasets. The pathway of the chain is finally analyzed to reconstruct the multidimensional posterior probability distribution and infer the sets of parameters that most likely reproduce the observed catalogs, as well as the correlations between these parameters. The main steps of this approach are detailed in the sections below, and the whole pipeline is sketched in \fg\ref{fig:workflow} of this article.


  \subsection{Physical parameters and source catalog generation}
\label{subsec:stuff}
Artificial catalogs are generated with the {\sc Stuff} package \citep{Bertin09} in fields of a given size. {\sc Stuff} relies on empirical scaling laws applied to a set of galaxy ``types'', which it uses to draw galaxy samples with photometric properties computed in an arbitrary number of observation passbands.
Each galaxy type is defined by its \citet{Schechter1976} luminosity function parameters, its spectral energy distribution (SED), as well as the bulge-to-total luminosity ratio $B/T$ and rest-frame extinction properties of each component of the galaxy through a ``reference'' passband.
 
The photometry of simulated galaxies is based on the composite SED templates of \citet{Coleman1980} extended by \citet{Arnouts1999}. Any of the six ``E'', ``S0'', ``Sab'', ``Sbc'', ``Scd'', and ``Irr'' SEDs can be assigned to the bulge and disk components separately, for a given galaxy type. The version of {\sc Stuff} used in this work does not allow the SEDs to evolve with redshift; instead, following \citet{Gabasch2004}, galaxy evolution is modeled as a combination of density (Schechter's $\phi^*$) and luminosity (Schechter's $M^*$) evolution with redshift $z$:
\begin{equation}
M^*(z) = M^*(0) + M_e\ln (1+z)
\end{equation}
\begin{equation}
\phi^*(z) = \phi^*(0) (1+z)^{\phi_e},
\end{equation}
where $M_e$ and $\phi_e$ are constants. The reference filter (\ie the filter where the LF is measured) is set to the g-band in the present article.
 
Bulges and elliptical galaxies have a \cite{DeVaucouleurs1953} profile:
\begin{equation}
      \mu_b(r) = M_b
                   + 8.3268 \left(\frac{r}{r_b}\right)^{\frac{1}{4}}
                   + 5 \log r_b + 16.6337,
\end{equation}
where $\mu_b(r)$ is the bulge surface brightness in ${\rm mag.pc}^{-2}$, $M_b = M - 2.5 \log (B/T)$ is the absolute magnitude of the bulge component and M the total absolute magnitude of the galaxy, both in the reference passband. As a projection of the fundamental plane, the average effective radius $\langle r_b \rangle$ in pc follows an empirical relation we derive from the measurements of \citet{Binggeli1984}):
\begin{equation}
      \langle r_b \rangle = \left\{
        \begin{array}{ll}
          r_{knee} 10^{- 0.3(M_b-M_{knee})} &
            \mbox{if $M_b < M_{knee}$} \\
          r_{knee} 10^{-0.1(M_b-M_{knee})} &
            \mbox{otherwise}
        \end{array}\right.
\end{equation}
where $r_{knee}=1.58 h^{-1}$kpc and $M_{knee}=-20.5$. The intrinsic flattening $q$ of bulges follows a normal distribution with $\langle q \rangle = 0.65$ and $\sigma_q = 0.18$ \citep{Sandage1970}, which we convert to the apparent aspect-ratio $\sqrt{q^2 \sin^2 i + \cos^2 i}$, where $i$ is the inclination of the galaxy with respect to the line of sight.
 
Disks have an exponential profile:
\begin{equation}
      \mu_d(r) = M_d + 1.8222 \left(\frac{r}{r_d}\right)
                       + 5 \log r_d + 0.8710,
\end{equation}
where  $\mu_d(r)$ is the disk surface brightness in ${\rm mag.pc}^{-2}$, $M_d = M - 2.5 \log (1 - (B/T))$ is the absolute magnitude of the disk in the reference passband, and $r_d$ the effective radius. Semi-analytical models where disks originate from the collapse of the baryonic content of dark-matter-dominated halos \citep{Dalcanton1997,Mo1998} predict useful scaling relations. Assuming that light traces mass and that there is negligible transport of angular momentum during collapse, one finds $r_d \propto \lambda L_d^{-\beta}$, where $\lambda$ is the dimensionless spin parameter of the halo, $L_d= 10^{-0.4 M_d}$ the total disk luminosity, and $\beta \simeq -1/3$ \citep{deJong2000}.
The distribution of $\lambda$, as seen in N-body simulations, can well be described by a log-normal distribution \citep{Warren1992}, and is very weakly dependent on cosmological parameters \citep{Steinmetz1995}, hence the distribution of $r_d$ at a given $M_d$ should behave as:
\begin{equation}
    n(r_d | M_d) \propto \frac{1}{r_d} \exp \left[-\frac{\left(\ln (r_d/r^*_d)
                        - 0.4\beta_d(M_d-M^*_d)\right)^2}{2\sigma^2_\lambda}\right].
\end{equation}
In \citet{deJong2000}, a convincing fit to I-band catalog data of late-type galaxies corrected for internal extinction is obtained, with $\beta_d=-0.214$, $\sigma_\lambda = 0.36$, $r^*_d = 5.93$~kpc, and $M^*_d = -22.3$ (for $H_0 = 65 {\rm km}.{\rm s}^{-1}$). Both bulge and disk effective radii are allowed to evolve (separately) with redshift $z$ using simple $(1+z)^\gamma$ scaling laws \citep[see, e.g.,][]{Trujillo2006,Williams2010}. The original values from \citet{Trujillo2006} are modified to those in Table~\ref{tab:sizeevolparams} based on the Hubble Space Telescope Ultra Deep Field (UDF, \citealt{Williams2010}, Bertin, private communication).
 
Internal extinction is applied (separately) to the bulge and disk SEDs $S(\lambda)$ using the extinction law from \cite{Calzetti1994}, extended to the UV and the IR assuming an LMC law (Charlot, private communication):
\begin{equation}
   S(\lambda) =  S_0(\lambda) e^{-\kappa\tau(\lambda)},
\end{equation}
where $S_0(\lambda)$ is the face-on, unextincted SED and $\tau(\lambda)$ the uncalibrated extinction law. The normalization factor $\kappa$ is computed by integrating the effect of extinction $A_{\rm ref}$, expressed in magnitudes, within the reference passband $p_{\rm ref}(\lambda)$:
\begin{equation}
A_{\rm ref} = -2.5 \log_{10} \frac{\int p_{\rm ref}(\lambda) S_0(\lambda) e^{ - \kappa \tau(\lambda)} d\lambda}{\int p_{\rm ref}(\lambda) S_0(\lambda) d\lambda}.
\end{equation}
 As the variation of $\tau(\lambda)$ is small within the reference passband, we take advantage of a second order Taylor expansion of both the exponential and the logarithm:
\begin{eqnarray}
\label{eq:quadraticext}
A_{\rm ref} & \approx & -2.5\log_{10} \left(1 - I_1\kappa + \frac{1}{2}I_2\kappa^2\right)\\
\label{eq:quadraticext2}
            & \approx &  1.086\left(I_1\kappa + \frac{I_1^2 - I_2}{2}\kappa^2\right),
\end{eqnarray}
with
\begin{equation}
I_1 = \frac{\int p_{\rm ref}(\lambda) S_0(\lambda) \tau(\lambda) d\lambda}{\int p_{\rm ref}(\lambda) S_0(\lambda) d\lambda},\ \ \ 
I_2 = \frac{\int p_{\rm ref}(\lambda) S_0(\lambda) \tau^2(\lambda) d\lambda}{\int p_{\rm ref}(\lambda) S_0(\lambda) d\lambda}.
\end{equation}
Solving the quadratic equation (\ref{eq:quadraticext2}) we obtain:
\begin{equation}
\kappa \approx \frac{-2A_{\rm ref}}{1.086\left(I_1 + \sqrt{I_1^2 - \frac{2}{1.086}(I_1^2 - I_2)A_{\rm ref}}\right)}.
\end{equation}
We adopt the parametrization of the extinction from the RC3 catalog \citep{DeVaucouleurs1991}:
\begin{equation}
    A_{\rm ref} = -\alpha(\text{T}) \log_{10} (\cos i),
\end{equation}
where $i$ is the disk inclination with respect to the line-of-sight, and $\alpha(\text{T})$ (not to be confused with Schechter's $\alpha$) is a type-dependent ``extinction coefficient'' that quantifies the amount of extinction+diffusion in the blue passband. For simplicity we identify this passband with our reference $g$ passband, although they do not exactly match. The extinction coefficient evolves with \citet{DeVaucouleurs59} revised morphological type as: 
\begin{equation}
\alpha(T) = \left\{
        \begin{array}{ll}
          1.5 - 0.03 (\text{T}-5)^2 &
            \mbox{for T $\geq$ 0} \\
          0 &
            \mbox{for T $\leq$ 0}
        \end{array}\right.
\end{equation}
 
{\sc Stuff} applies to SEDs the mean intergalactic extinction curve at the given redshift following \cite{Madau1995} and \cite{Madau1996}, using the list of Lyman wavelengths and absorption coefficients from the {\sc Xspec} code \citep{Arnaud1996}. Galaxies are Poisson distributed in $5 h^{-1}$~Mpc redshift slices from $z=20$ to $z=0$. For now the model does not include clustering properties, therefore the galaxies positions are uniformly distributed over the field of view. Ultimately {\sc Stuff} generates a set of mock catalogs (one per filter) to be read by the image simulation software, containing source position, apparent magnitude, $B/T$, bulge and disk axis ratios and position angles, and redshift. We note that for consistency, we kept most of the default values applied by {\sc Stuff} to scaling parameters, although many of them come from slightly outdated observational constraints dating back to the mid-2000's (and even earlier). This of course does not affect the conclusions of this paper.
 
  \subsection{Image generation}
  \label{subsec:skymaker}
\begin{figure*}
  \includegraphics[width=\textwidth]{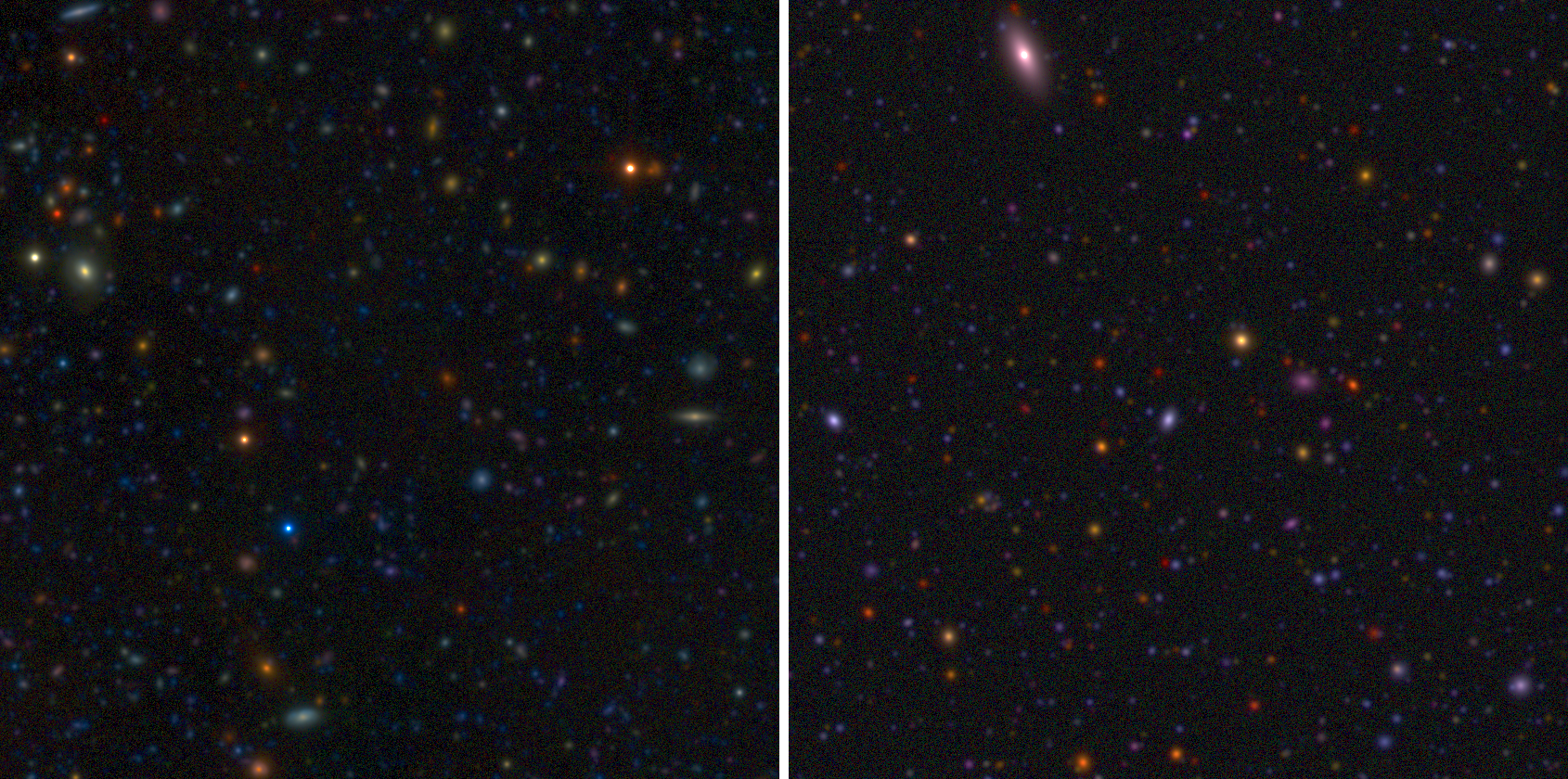}
  \caption{Comparison between an observed survey image and a mock image generated by our model. On the left: a region of the CFHTLS D1 field (stack from the $85\%$ best seeing exposures) built from the \textit{gri} bands. On the right: a simulated image with {\sc Stuff}+{\sc Skymaker} with the same filters, exposure time, and telescope properties as the CFHTLS data. Both images are shown with the same color coding.}
  \label{fig:imexample}
\end{figure*}

  {\sc Stuff} catalogs are turned into images using the {\sc SkyMaker} package \citep{Bertin09}. Briefly, {\sc SkyMaker} renders simplified images of galaxy models as the sum of a \cite{Sersic1963} ``bulge'' and an exponential ``disk'' on a high resolution pixel grid. The models are convolved by a realistic PSF model generated internally, or derived from actual observations using the {\sc PSFEx} tool \citep{Bertin2011}. Each convolved galaxy image --- or point source for stars --- is subsampled at the final image resolution using a Lanczos-3 kernel \citep{Wolberg1990} and placed on the pixel grid at its exact catalog coordinates. The next step involves large scale features: convolution by a PSF aureole \citep[e.g.,][]{Racine1996}, addition of the sky background, and simulation of saturation features (bleed trails). Finally, photon (Poisson) and read-out (Gaussian) noise are added according to the characteristics of the instrument being simulated, and the data are converted to ADUs (analog-to-digital units). An example of a simulated deep survey field is shown Fig. \ref{fig:imexample}.
 
\section{Compression of data: from source extraction to binning}
 
\label{sec:source_extract}
 
  \subsection{Source extraction}
  \label{subsec:sextractor}
 The {\sc SExtractor} package \citep{Bertin1996} produces photometric catalogs from astronomical images. 
  Briefly, sources are detected in four main steps: first, a smooth model of the image background is computed and subtracted. Second, a convolution mask, acting as matched filter, is applied to the background-subtracted image for improving the detection of faint sources. Third, a segmentation algorithm identifies connected regions of pixels with a surface brightness in the filtered image higher than the detection threshold. Finally, the same segmentation process is repeated at increasing threshold levels to separate partially blended sources that may share light at the lowest level.
 
  Once a source has been detected, {\sc SExtractor} performs a series of measurements according to a user-defined parameter list. This includes various position, shape, and flux estimates. For this work we rely on \textbf{FLUX\_AUTO} photometry. \textbf{FLUX\_AUTO} is based on Kron's algorithm \citep{Kron1980} and gives reasonably robust photometric estimates for all types of galaxies. For object sizes we choose the half-light radius estimation provided by the \textbf{FLUX\_RADIUS} parameter, which is the radius of the aperture that encloses half of the \textbf{FLUX\_AUTO} source flux. We note that this size estimate includes the convolution of the galaxy light profile by the PSF.
In order to retrieve properties such as color, {\sc SExtractor} is run in the so-called double image mode, where detection is carried out in one image and measurements in another. By repeating source extraction with the same ``detection image'', but with ``measurement images'' in different filters, we ensure that the photometry is performed in the exact same object footprints in all filters.
 
{\sc SEXtractor} flags all issues occurring during the detection and measurements processes. In this work, we consider only detections with a {\sc SExtractor} \textbf{FLAG} parameter less than four, which excludes sources that are saturated or truncated by the frame boundaries.
 
\subsection{Parallelization}
\label{parallel}

\begin{figure*}
  \includegraphics[width=\textwidth]{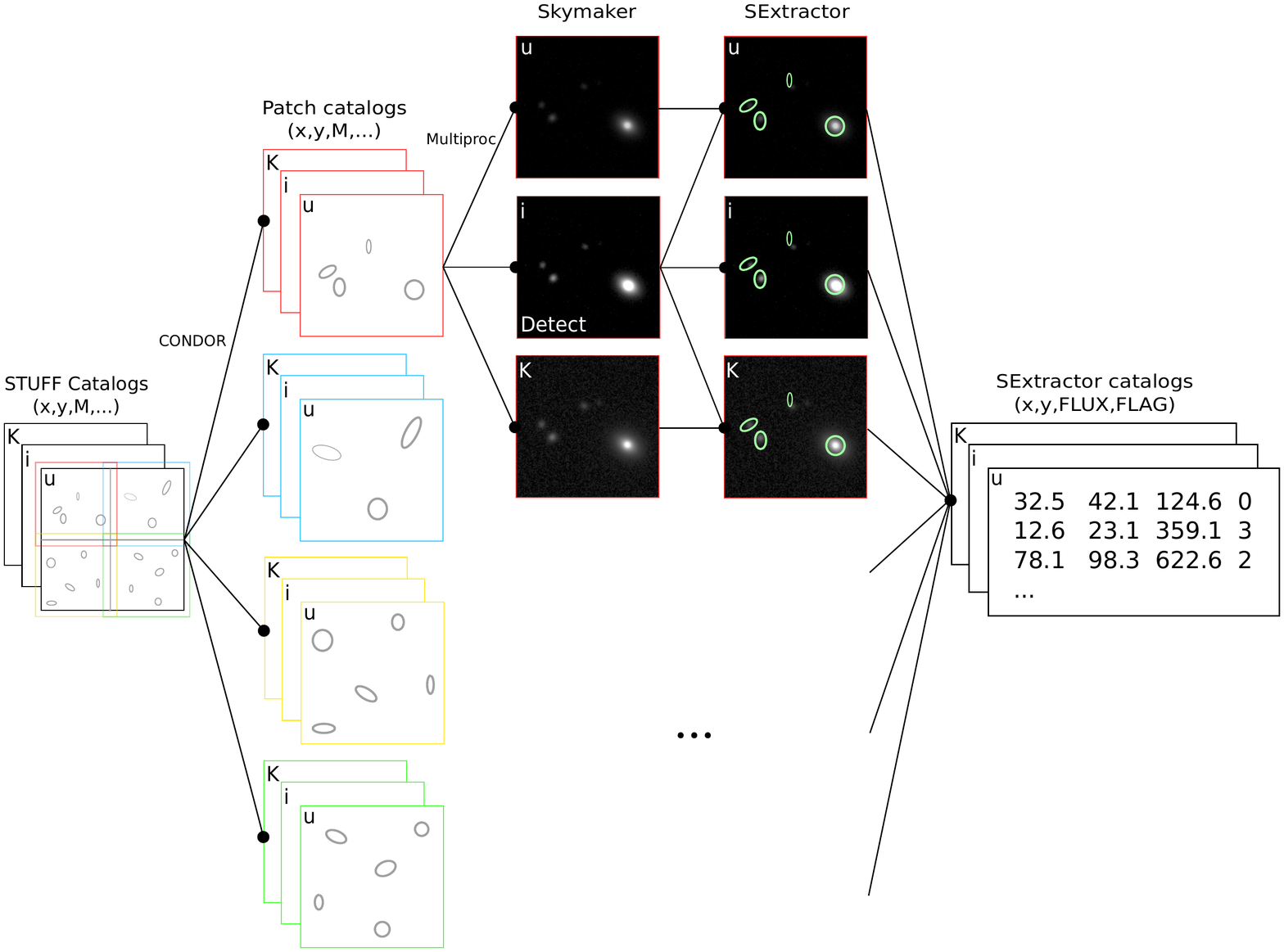}
  \caption{Illustration of the parallelization process of our pipeline, described in detail in section \ref{parallel}. {\sc Stuff} generates a catalog, that is, a set of files containing the properties of simulated galaxies, such as inclination, bulge-to-disk ratio, apparent size, and luminosity. Each file lists the same galaxies in a different passband. The parallelization process is performed on two levels: first, the {\sc Stuff} catalogs are split into sub-catalogs according to the positions of the sources on the image. These sub-catalogs are sent to the nodes of the computer cluster in all filters at the same time using the HTCondor framework. Each sub-catalog is then used to generate a multiband image corresponding to a fraction of the total field. This step is multiprocessed in order to generate the patches in every band simultaneously. {\sc SExtractor} is then launched on every patch synchronously, also using multiprocessing. The source detection is done in one pre-defined band, and the photometry is done in every band. Finally, the {\sc SExtractor} catalogs generated from all the patches are merged into one large catalog containing the photometric and size parameters of the extracted sources from the entire field.} 
  \label{fig:split}
\end{figure*}

\begin{figure*}
  \includegraphics[width=\textwidth]{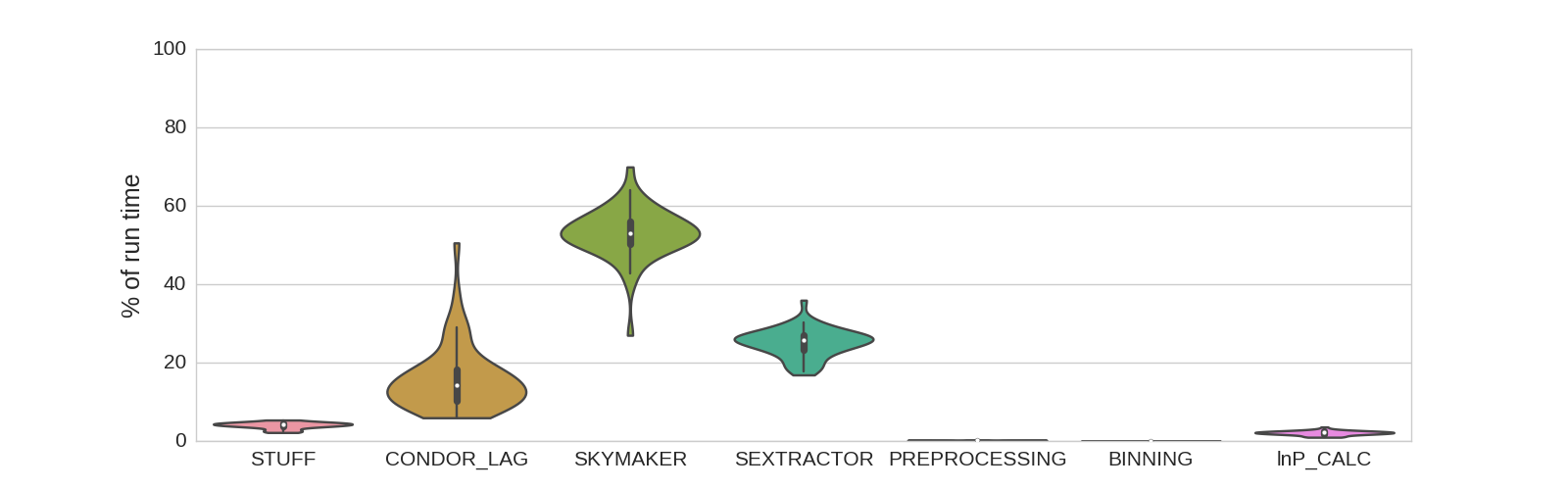}
  \caption{Benchmarking of a full iteration of our pipeline, obtained with 50 realizations of the same iteration. An iteration starts with the {\sc Stuff} catalog generation (here we consider a case where $\sim$ 55000 sources spread into two populations of galaxies are produced), and ends with the posterior density computation. The runtime of each subroutine called is analyzed in terms of the fraction of the total runtime of the iteration. In this scheme, the image simulation step clearly dominates the runtime, followed by the source extraction step and the HTCondor latencies. Source generation, pre-processing, binning and posterior density calculation (labeled lnP\_CALC), however, account for a negligible fraction of the total runtime.}
   \label{fig:bench}
\end{figure*}

By construction, our sampling procedure based on MCMC (cf section \ref{subsec:mcmc}) cannot be parallelized, because the knowledge of the $n-1_{th}$ iteration is required to compute the $n_{th}$ iteration. We can, however, parallelize the process of source extraction and, most importantly, image simulation. In fact, we find in performance tests that the pipeline runtime is largely dominated by the image generation process  (cf \fg\ref{fig:bench}), and that the image generation time scales linearly with the area of the simulated image. Simulating a single image per band containing all the sources for every iteration would make this problem computationally unfeasible in terms of execution time. In order to limit the runtime of an iteration, the image making step is therefore split into $N_{sub} \times N_{f}$ parallel small square patches, as illustrated in \fg\ref{fig:split}, where $N_{f}$ is the number of filters fixed by the observed data and $N_{sub}$ the user-defined number of patches per band. Both quantities must be chosen so that their product optimizes the resources used by the computing cluster.
 
We start with $N_{f}$ input catalogs generated from the model, each containing a list of sources' positions in a full-sized square field of size $L_f$, as well as their photometric and size properties. The sources are then filtered according to their spatial coordinates and dispatched to their corresponding patch. Each patch has a size $L_f/\sqrt{N_{sub}}$, where $N_{sub}$ is a square number. In practice, the sources are extracted from a box 150 pixels wider than the patch size in order to include the objects outside the frame that partially affect the simulated image. 
All the sources of position (x,y) are within a patch of coordinate $(i,j)$ $\in [0,\sqrt{N_{sub}}-1] \times [0,\sqrt{N_{sub}}-1]$ if
$x \in [i \frac{L_f}{\sqrt{N_{sub}}}-150, (i+1) \frac{L_f}{\sqrt{N_{sub}}}+150]$, and 
$y \in [j \frac{L_f}{\sqrt{N_{sub}}}-150, (j+1) \frac{L_f}{\sqrt{N_{sub}}}+150]$.
 
As a result, all the sources are scattered through $N_{sub}$ catalog files per band. We then use the HTCondor distributed jobs scheduler on our computing cluster to generate and analyze all the patches at the same time. The flexibility of HTCondor offers many advantages to a pipeline that requires distributed computing over long periods of time. Thanks to its dynamic framework, jobs can be check pointed and resumed after being migrated if a node of the cluster becomes unavailable, and the scheduler efficiently provides an efficient match-making between the required and the available resources. This framework also has its drawbacks, in the form of inherent and uncontrollable latencies when jobs input files are sent to the various nodes.
 
In our case, each job corresponds to a single patch, and the $N_{sub} \times N_{f}$ resulting catalogs serve as input files for the jobs. We found that HTCondor latencies represent between 7\% and 50\% of the run time of each iteration, as illustrated in \fg\ref{fig:bench} in the context of the application described below (cf \sct\ref{sec:application}).
 
For each job, the image generation and source extraction procedures are multiprocessed: {\sc SkyMaker} is first launched simultaneously in every band on the $L_f/\sqrt{N_{sub}}$-sized patch and, when all the images are available, {\sc SExtractor} is launched in double image mode. Condor then waits until all jobs are completed. Finally, the catalog files generated from all the patches are merged into one, so that \textit{in fine}, a single catalog file per band contains all the extracted sources.
 
\subsection{Reduction of the dynamic ranges}
\label{subsec:reddynrange}
 
Observables such as fluxes may have a large dynamic range that goes up to the saturation level of the chosen survey. This can be problematic for the binning process of our pipeline, in the sense that it will create many sparsely populated bins. We must therefore reduce the dynamic range of the photometric properties of the sources. We cannot simply use the log of the flux arrays, because the noise properties of background-subtracted images can provide faint objects with negative fluxes. We therefore use the following transform $g(X)$, which has already been applied to model-fitting and machine learning applications (e.g., \citealt{2011ascl.soft10006B}):
 
\begin{align}
X_r = g(X) = \left\{\begin{array}{rl}
  \kappa_{c}\sigma \ln \left(1 + \frac{X}{\kappa_{c}\sigma}\right) & \mbox{if $X\ge 0$,}\\
 	-\kappa_{c}\sigma \ln \left(1 - \frac{X}{\kappa_{c}\sigma}\right) & \mbox{otherwise,}\\
 \end{array}\right.
\label{eq:reducrange}
\end{align}
where $\sigma$ is the baseline standard deviation of $X$ (\ie, the average lowest flux error), and $\kappa_{c}$ a user-defined factor which can be chosen in the range from 1 to 100, typically. In all the test cases that we describe in \sct\ref{sec:application}, we set $\kappa_{c}=10$. In practice we apply this compression to each dimension of the observable space, with a different value of $\sigma$ for each observable. We separate the $\sigma$ values into two categories for each kind of observable: $\sigma_{f}$ for flux-related observables and $\sigma_{r}$ for size-related ones. These values are affected by the galaxy populations in the observed field as well as the photometric properties of the field itself, such as the bands used and the noise properties. For fluxes and colors, a root mean square error estimate of the flux measurement is given by {\sc SExtractor}: \textbf{FLUXERR\_AUTO}. We set $\sigma_{f}$ to the median value of the distribution of \textbf{FLUXERR\_AUTO} values for the sources extracted from input data, and this operation is repeated on each filter. However, {\sc SExtractor} provides no such error estimate for \textbf{FLUX\_RADIUS}. For this kind of observable we rely on the distribution of \textbf{FLUX\_RADIUS} of the extracted sources with respect to the corresponding \textbf{FLUX\_AUTO}. For each passband, the value of $\sigma_{r}$ is set to the approximate \textbf{FLUX\_RADIUS} of the extracted sources' distribution when \textbf{FLUX\_AUTO} tends to 0. The exact values actually do not matter, because the same compression is applied on the observed and simulated data.
 
\subsection{Decorrelation of the observables: whitening transformation}
\label{subsec:whitening}
 
The choice of the nature and number of observables is a compromise between computational cost and informational content. In fact, memory limitations intrinsic to the computational cluster when binning observed and synthetic data (cf \sct\ref{subsec:binning}) prevent us from using an arbitrary number of observables in the pipeline. Observables such as fluxes or magnitudes in different passbands also tend to be correlated with one another, as they originate from the same spectrum of a given galaxy from a given population. These correlations can be high if the passbands are too narrow, too close to each other, and not covering a large enough wavelength baseline. One must thoughtfully choose the appropriate set of filters \textit{a priori} in order for the resulting set of observables to be able to disentangle the luminous properties of the different galaxy populations.
 
Strong correlations between input vector components can also make binning very inefficient, therefore an important pre-processing step is to decorrelate them. In that regard, we apply a linear transformation called principal component analysis whitening, or sphering (\citealt{Friedman1987}, \citealt{Hyvarinen2009}, \citealt{Shlens2014}, \citealt{Kessy2016}) to our reduced matrix of observables $X_r$ of size $p \times N_{s}$, where $p$ is the number of observables and $N_{s}$ is the number of sources. Principal component analysis (PCA) is an algorithm commonly used in the context of dimensionality reduction. Its goal is to find a set of orthogonal axes in a dataset called principal components that encapsulate most of the variance of the data. This can be performed via a singular values decomposition (SVD) of the covariance matrix of the data:
 
\begin{align}
<X_rX_r^T>=U\Lambda V^T,
\end{align}
where U and V are orthogonal matrices and $\Lambda$ the diagonal matrix containing the non-negative singular values of the covariance matrix, sorted by descending order.
 
PCA whitening is the combination of two operations: rotation and scaling. First the dataset (previously centered around zero by subtracting the mean in each dimension) is projected along the principal components, which removes linear correlations, and then each dimension is scaled so that its variance equals to one. The whitening transform can therefore be summarized by:
 
\begin{align}
X_w=\Lambda^{-\frac{1}{2}}V^T(X_r-\mu),
\end{align}
where $X_w$ is the whitened version of the observables matrix $X_r$ and $\mu$ is the average matrix.
The PCA whitening transformation results in a set of new variables that are uncorrelated and have unit variance ($<X_wX_w^T>=I$). During the chain iterations, the observed and simulated data are centered, rotated, and scaled in the same way to ensure that both distributions can be well superposed and compared (cf section~\ref{sec:comparison}). %
 
In practice, the simulated data is whitened using the $\Lambda$, $V^T$, and $\mu$ of the observed data. The number of principal components to keep is left to the choice of the user. Retaining only the components with the highest variance and therefore reducing the computational cost of the pipeline may be tempting. Nevertheless, subtle but important features can arise from low variance components, and deleting them comes at a price. In our application (cf section~\ref{sec:application}), we choose not to reduce the dimensionality of the problem.
 
\begin{figure*}
  \begin{subfigure}{0.45\textwidth}
    \centering
    \includegraphics[width=\textwidth]{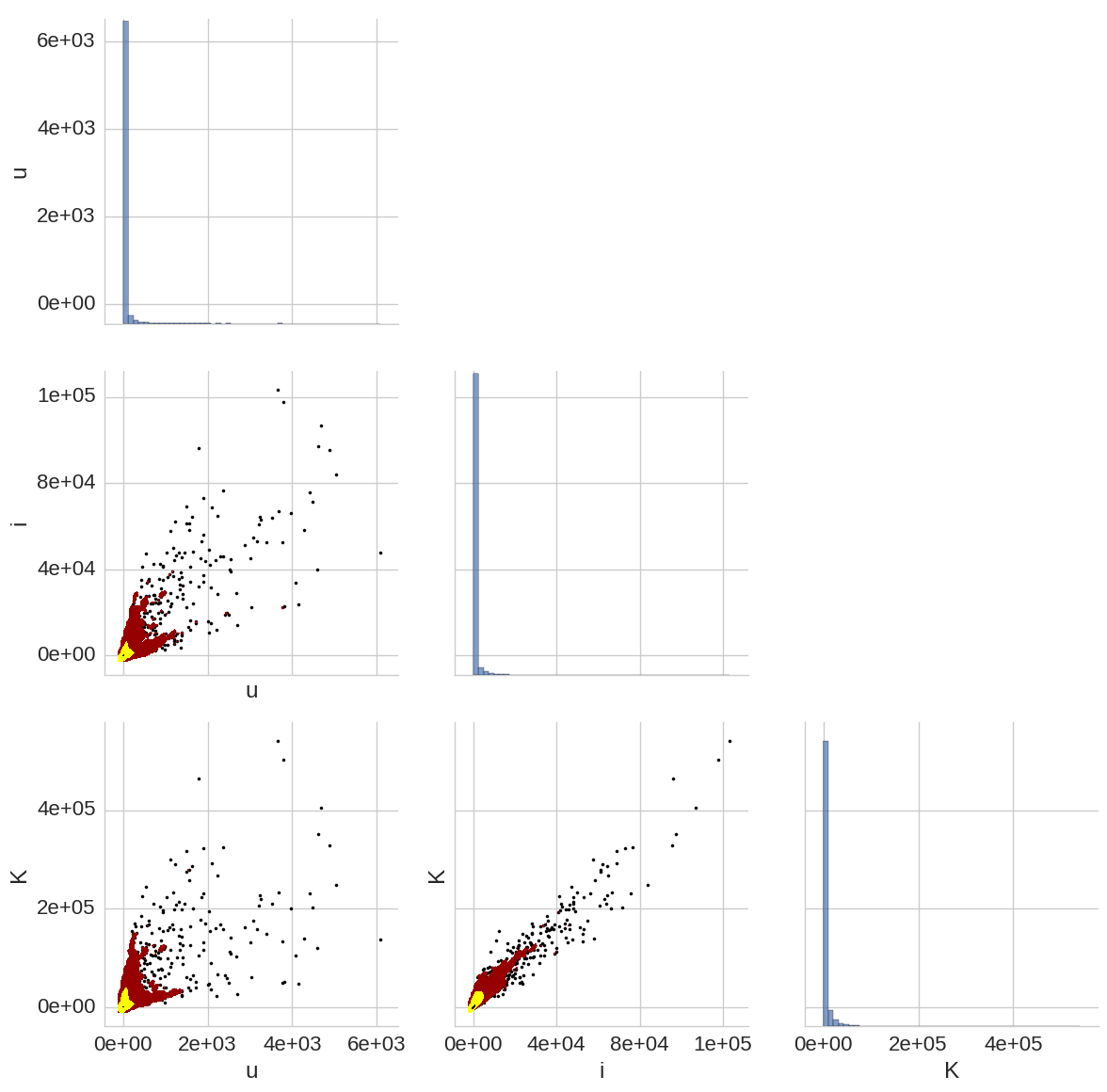} 
    \caption{Initial} 
    \label{bla0} 
    \vspace{4ex}
  \end{subfigure}
  \begin{subfigure}{0.45\textwidth}
    \centering
    \includegraphics[width=\textwidth]{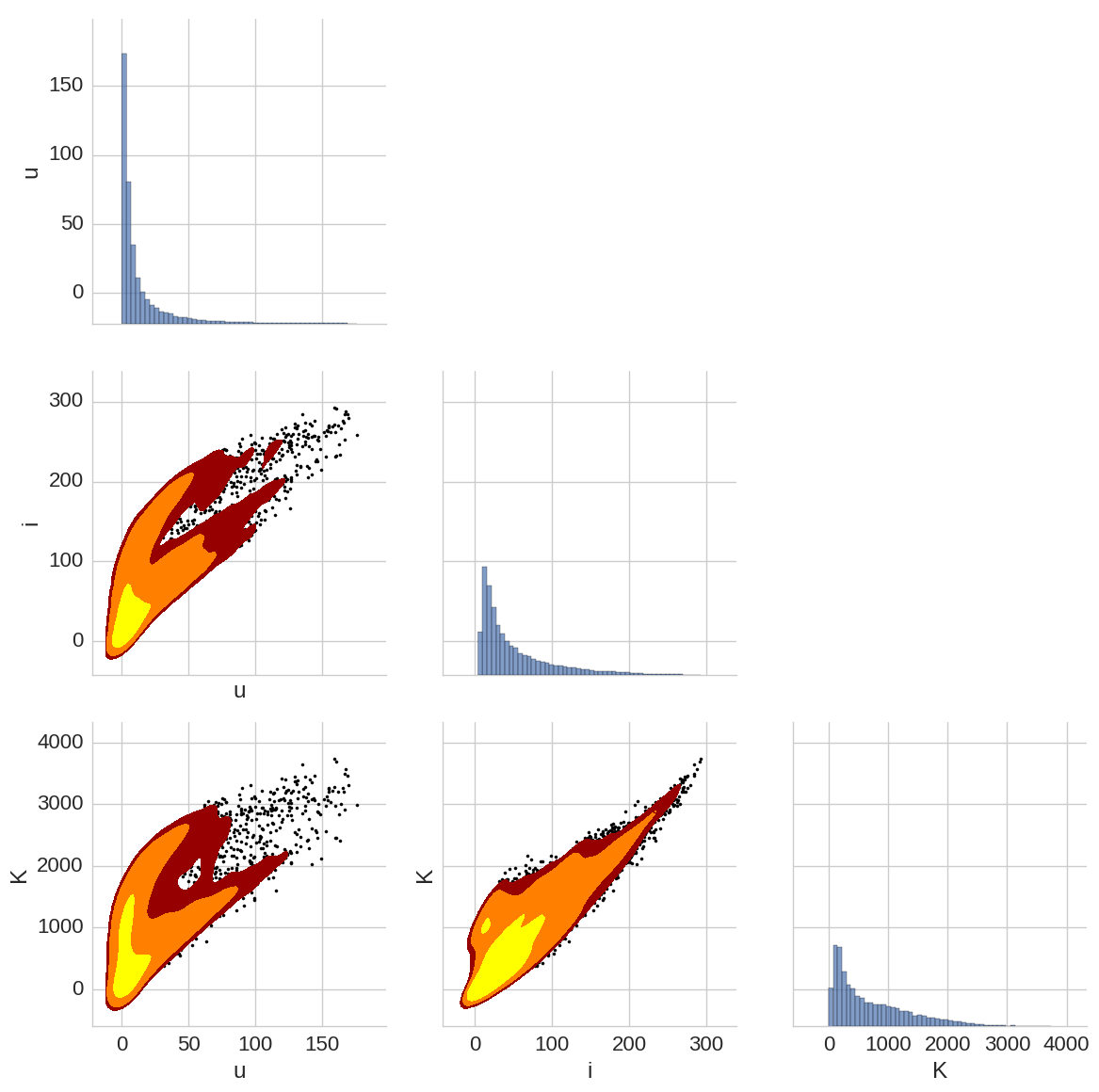} 
    \caption{Reduced} 
    \label{bla1} 
    \vspace{4ex}
  \end{subfigure} 
  \centering
  \begin{subfigure}{0.45\textwidth}
    \centering
    \includegraphics[width=\textwidth]{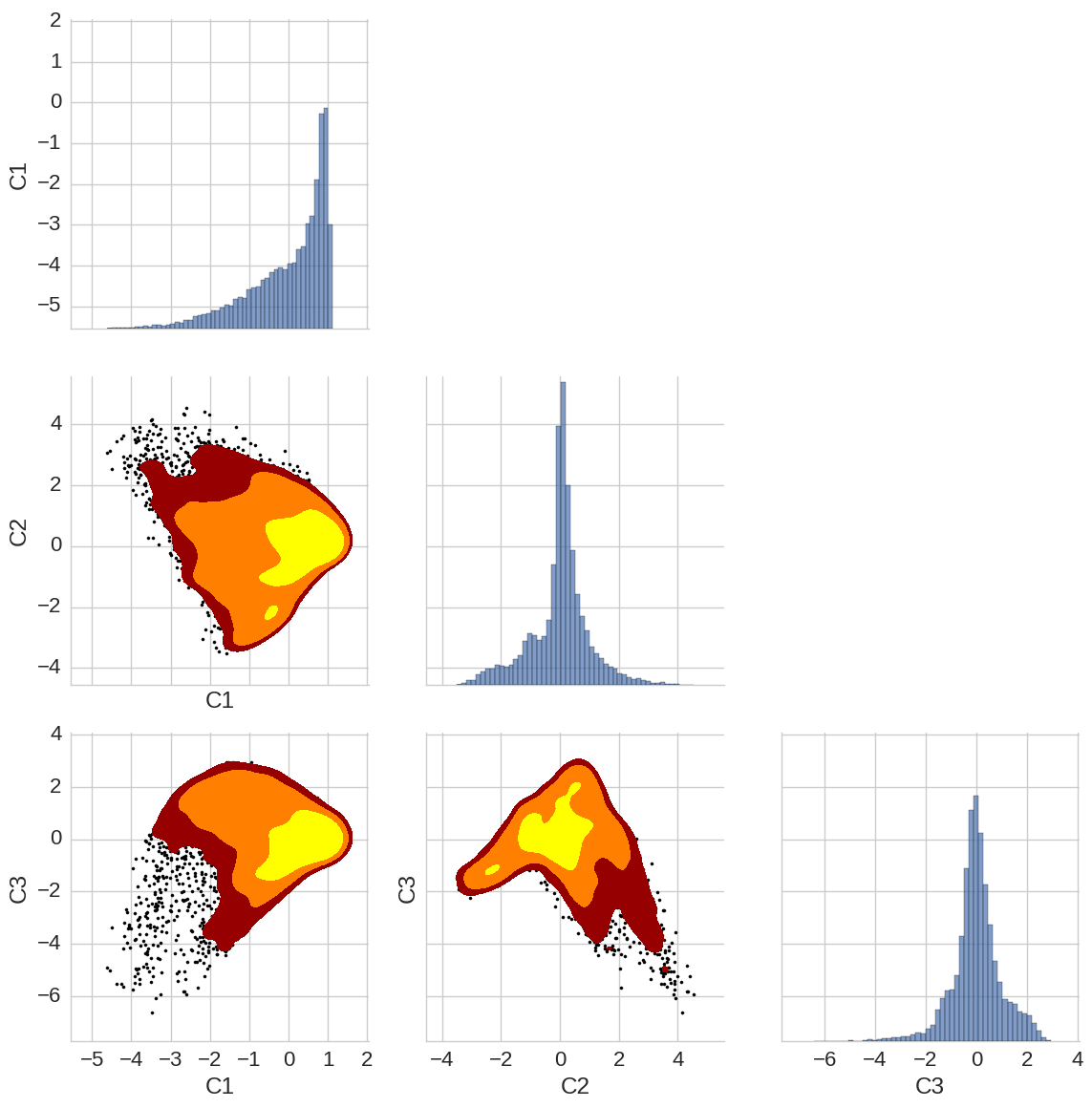} 
    \caption{Whitened} 
    \label{bla2} 
  \end{subfigure} 
  \caption{Distribution of observables before and after each step of pre-processing from the mock input data with 2 populations of galaxies (Ellipticals+Spirals) described in \sct\ref{subsec:2_pop}. The dark red, orange and yellow areas in the contour plots are the regions that enclose 99\%, 95\% and 68\% of the points respectively. \textit{Top left panel}: scatter plot of \textbf{the FLUX\_AUTO} of extracted sources (in ADUs) in filters \textit{ui$K_s$} and their covariances. \textit{Top right panel}: same plot, but with the dynamic range of the \textbf{FLUX\_AUTO} distributions reduced via \eq \ref{eq:reducrange}. \textit{Bottom panel}:  same plot, after whitening of the reduced distribution of observables. The latter distribution is uncorrelated, centered on the mean of the distribution and rescaled, allowing for a much more efficient binning process than on raw fluxes, and a more practical comparison with the simulated observables.}
  \label{bla3} 
\end{figure*}
 
\begin{figure*}
  \begin{subfigure}{0.45\textwidth}
    \centering
    \includegraphics[width=\textwidth]{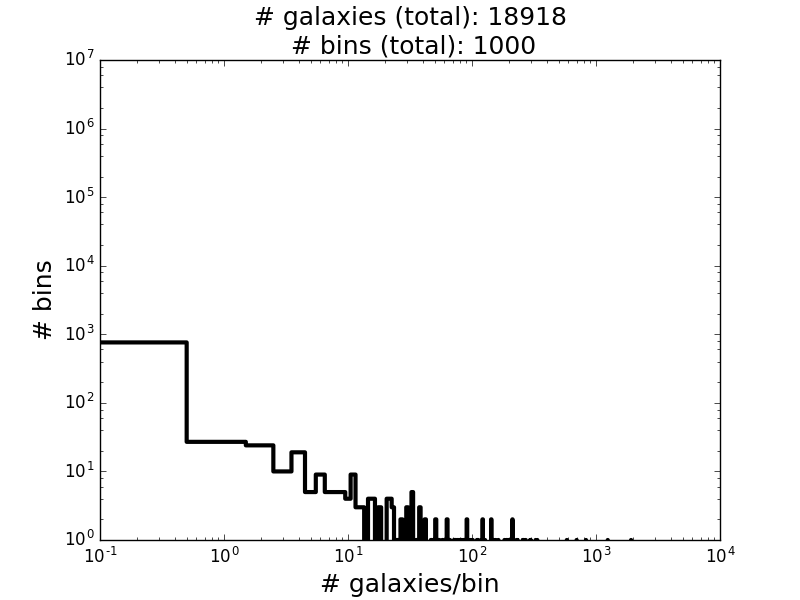} 
    \caption{Multi-type case} 
    \label{histbins2pop} 
    \vspace{4ex}
  \end{subfigure}
  \begin{subfigure}{0.45\textwidth}
    \centering
    \includegraphics[width=\textwidth]{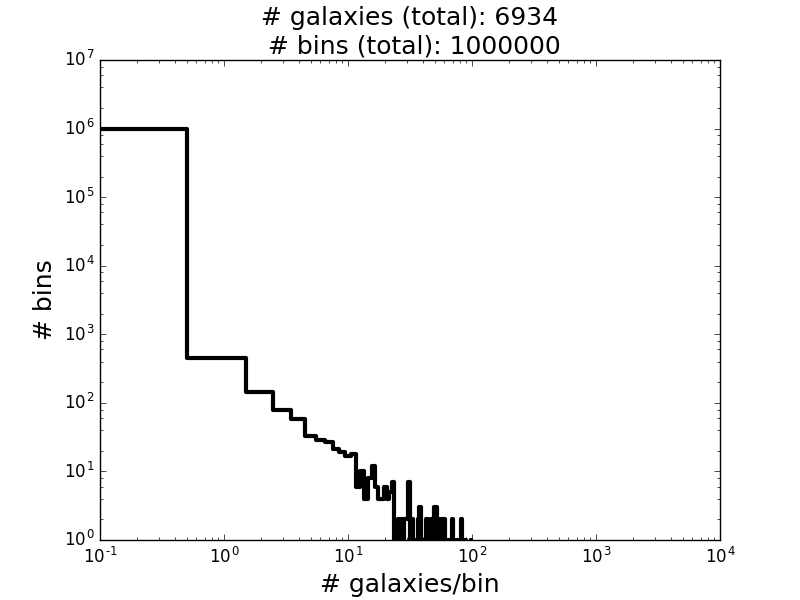} 
    \caption{Fattening E case} 
    \label{histbins1pop} 
    \vspace{4ex}
  \end{subfigure} 
  \caption{Histogram of the number of sources extracted per bin for the pre-processed input data of the test cases presented in \sct\ref{sec:application}. In the left panel, three observables are considered: the \textbf{FLUX\_AUTO} in \textit{ui$K_s$}. In the right panel, six observables are considered: the \textbf{FLUX\_AUTO} in \textit{ui$K_s$} and the \textbf{FLUX\_RADIUS} in \textit{ui$K_s$}. With the binning rule described in \sct\ref{subsec:binning}, between the ``Multi-type'' case and the ``Fattening E'' case, the number of bins increases by a factor $10^3$, and the number of empty bins is increased by roughly the same amount. This illustrates the curse of dimensionality we face in this method, and puts computational limits on the number of observables we can use.}
  \label{fig:histbins}
  \end{figure*}
 
\subsection{Binning in the observational space}
\label{subsec:binning}
 
It remains to quantify the similarity between the two multivariate datasets, one containing preprocessed observables from the observations and the other from a simulation. Following the idea of \citet{Robinetal14} and \citet{Rybizki2015}, who grouped their data representing stellar photometry into bins of magnitude and color, we choose to bin our datasets, considering the relative simplicity and advantageous computational cost of this method. However, binning comes with some inevitable drawbacks: the number of bins increases exponentially with the number of dimensions. For a fixed-size dataset, multivariate histograms are also sparser than their univariate counterparts and display more complex shapes. Finally the choice of the binning scheme can significantly influence the information content of the dataset, and that choice is not trivial in high-dimensional spaces (\citealt{Cadez2002}). This class of problems is known as ``the curse of dimensionality'' (\citealt{Bellman1972}). 
 
Several binning schemes have been developed, like the  \citet{Freedman1981} rule extended to several dimensions, Knuth's rule (\citealt{Knuth2006}), which uses Bayesian model selection to find the optimal number of bins, Hogg's rule (\citealt{Hogg2008}), or Bayesian blocks (\citealt{Scargle2012}). But all these rules face the curse of dimensionality as the number of observables becomes high. Alternatives to binning for density estimation can also be used and are discussed in \sct\ref{sec:discussion}.
 
In our specific case, the dimensionality of the observable space is determined by the number $p$ of photometric and size parameters in every passband extracted from the survey images. We use ten bins of constant width per dimension throughout the article. More bins per dimension would lead to memory issues caused by the limitations of our computing cluster in the applications that we propose in \sct\ref{subsec:1popE}. The bin width for dimension $k \in [1,p]$ in this scheme is therefore given by: 
 \begin{align}
W_k = \frac{\text{max}(X_{w,k}) - \text{min}(X_{w,k})}{10},
\end{align}
where $X_{w,k}$ is the pre-processed observables matrix for the observed data. 

In this pipeline, the binning pattern is only computed once and for the observed data only. The same binning is then directly applied to the simulated data to ensure better execution speed and comparability between histograms. Because the number of counts per bin is directly affected by the model parameters that rule the number density of galaxies, such as $\phi^*$ in our application (see \sct\ref{sec:application}), the resulting $p$-dimensional histograms are not normalized to prevent a loss of information in the minimization of distance between the synthetic and observed data.
 
 
\section{Comparison between simulated and observed data}
\label{sec:comparison}
 
Estimating the discrepancy between the observed and simulated binned datasets in high-dimensional space is highly non-trivial, as the choice of a good distance metric is problem dependent. The observables' distributions may be multimodal and skewed, and many metrics rely on the assumption of normality. Others, such as the Kullback-Leibler divergence (\citealt{Kullback1951}) or  the Jensen-Shannon distance (\citealt{Lin1991}), cannot be used without estimating an analytical underlying PDF, which can be very computationally expensive in a high-dimensional observable space.
 
Here is a non-exhaustive list of non-parametric (\ie, distribution-free) distance metrics found in the literature that can be used on multivariate data in the ABC framework. A more complete review is available in \citet{Pardo2006} and \citet{Palombo2011}; however, no study to quantify their relative power has been performed so far. These metrics include:
 
\begin{itemize}
\item the \textbf{$\chi^2$ test} (\citealt{Chardy1976}) is a simple and widely used way of determining whether observed frequencies are significantly different from expected frequencies. The main drawback of this approach is that $\chi^2$ test results are  dependent on the binning choice (\citealt{Aslan2002}). For example, \citet{Kurinsky2014} use the $\chi^2$ distance to compare color-color histograms.
\item the \textbf{Kolmogorov-Smirnov (KS) test}  (\citealt{chakravarti67}) estimates the maximum absolute difference between the empirical distribution functions (EDF) of two samples. A generalization of this test for multivariate data has been proposed (\citealt{Justel1997}). However, as there is no unique way of ordering data points to compute a distance between two EDF, it is not as reliable as the 1one-dimensional version without the help of resampling methods such as bootstraping \citep{babu06}.
\item the \textbf{Anderson-Darling (AD) test} (\citealt{stephens74}) is a modification of the KS test. This method uses a weight function that gives more weight to the tails of the distributions. It is therefore considered more sensitive than the KS test, but it also suffers from the same problems in the multivariate case.
\item the \textbf{Mahalanobis distance} (\citealt{citeulike:4155812}) is similar to the Euclidean norm but has the advantage of taking into account  the correlation structure of multivariate data. The Mahalanobis statistics, coupled with an univariate KS test, are used by \citet{Akeretetal15} to compare photometric parameters for cosmological purposes. However, this distance only works for unimodal data distributions.
\item the \textbf{Bhattacharyya distance} (\citealt{10.2307/25047882}) is related to the Bhattacharyya coefficient, which measures the quantity of overlap between the two samples. It is considered more reliable than the Mahalanobis distance in the sense that its use is not limited to cases where the standard deviations of the distributions are identical.
\item the \textbf{Earth Mover's distance (EMD)} (\citealt{Rubner}) is based on a solution to the Transportation problem. The distributions are represented by a user-defined set of clusters called signatures, where each cluster is described by its mean and by the fraction of the distribution encapsulated by it. The EMD is defined as the minimum cost of turning one signature into the other, the cost being linked to the distance between the two. A computationally fast approximate version of this distance using the Hilbert space-filling curve can be found in \citet{Bernton2017}.
\end{itemize}
 
In the present article, we place ourselves within the pBIL framework to perform the inference process. In this context, the binning structure constructed in section \ref{subsec:binning} and the assumption of a Poisson behavior of the number counts in each bin represent the auxiliary model that describes the data. The ``auxiliary likelihood''  derived from this structure is inspired from the maximum likelihood scheme of \citet{Cash1979}, a likelihood that has been used in previous studies like \citet{Robinetal14}, \citet{Bienayme1987}, or \citet{adyephd98}:
\begin{align}
\ln L = \sum_{i=1}^{b} (o_i \ln(s_i)-s_i )
\label{lnL}
\end{align}
where b is the total number of bins,  $s_i$ is the number count in bin $i$ for the simulated data, and $o_i$ is the number count in bin i for the observed data. The underlying assumptions for this choice of auxiliary likelihood can be found in Appendix \ref{app:demolnL}. \\

In that scheme, as the logarithm of $s_i$ is used, empty bins cause a problem. In order to avoid  singularities, a constant small value (that
 we set to 1) is added to every bin up to the edges of the observables space. This process is done in both modeled and observed data so that it does not bias our results.
 
\section{Sampling procedure: Adaptive Proposal algorithm} 
\label{subsec:mcmc}
 
\begin{algorithm}
 
\textbf{Initialize} parameters $\theta_{(0)}$ from prior distribution \;
\textbf{Initialize} covariance matrix and temperature \;
\For{t = 0 to $T$}
    {\textbf{Every} $S$ iterations: \\
    \qquad Update covariance matrix and temperature\;
    Propose new state $\theta^*$ from proposal distribution\;
    \While {$\theta^*$ is outside the prior bounds}{Propose another state}
 
  Compute $\ln P(\theta^*|\mathcal{D})$ from proposed state (\eq \ref{eq:bayes2}) \\
    \eIf {$\ln P(\theta^*|\mathcal{D})$ $\geq$ $\ln P(\theta_{(t)}|\mathcal{D})$}{Accept the jump}{Compute acceptance probability $a$ \;
    Draw uniformly distributed random number $R_N$ in the interval $[0,1]$ \;
    \eIf {$R_N < a$}{Accept the jump}{Refuse the jump}}
  }
\caption{Proposed sampling algorithm based on the AP algorithm \citep{Haario1999}.}
\label{alg:sampling}
\end{algorithm}
 
MCMC methods are a set of iterative processes which perform a random walk in the parameter space to approximate the posterior distribution with the help of Markov chains. A Markov chain is a sequence of random variables $\{\theta_{(0)},\theta_{(1)},\theta_{(2)},...,\}$ in the parameter space (called states) that verifies the Markov property: the conditional distribution of $\theta_{(t+1)}$ given $\{\theta_{(0)},...,\theta_{(t)}\}$ (called transition probability or kernel) only depends  on $\theta_{(t)}$. In other words, the probability distribution of the next state only depends on the current state. 
 
After a period (whose length depends on the starting point and the random path taken by the chain) where the chain travels from low to high probability regions of the parameter space, the MCMC samples ultimately converge to a stationary distribution in such a way that the density of samples is proportional to the posterior PDF, also called target distribution. The portion of the chain which is not representative of the target distribution (\ie, the first iterations where the chain has not yet reached stationarity) is called burn-in, and is usually discarded from the analysis a posteriori. Well optimized MCMC methods provide an efficient tool to avoid wasting a lot of computing time sampling regions of very low probability. There is a great variety of MCMC algorithms, and the choice of a specific algorithm is problem-dependent. The reader is referred to \citet{Roberts2009} for a complete review of these methods.
 
To estimate the posterior distribution $P(\theta|D)$ defined in \eq \ref{eq:bayes} in a reasonable amount of time, one must explore the parameter space in a fast and efficient way. For our purposes, we designed a custom sampling procedure, described in Algorithm \ref{alg:sampling}, based on the MCMC Adaptive Proposal (AP) algorithm (\citealt{Haario1999}), which is itself built upon the Metropolis-Hastings algorithm (\citealt{Metropolis1953}, \citealt{HASTINGS1970}). The Metropolis-Hastings algorithm is one of the most general MCMC methods. In this algorithm, given a state $\theta_{(t)}$ sampled from the target distribution $P(\theta)$, a proposed state $\theta^*$ is generated using a user-defined transition kernel $Q(\theta^*|\theta_{(t)})$, which represents the probability of moving from $\theta_{(t)}$ to $\theta^*$. The proposition is accepted with probability:
\begin{equation}
a=min\left\{ \frac{P(\theta^*)}{P(\theta_{(t)})} \frac{Q(\theta_{(t)}|\theta^*)}{Q(\theta^*|\theta_{(t)})},1 \right\}.
\label{accratemh}
\end{equation}
If the proposed sample is accepted, then $\theta_{(t+1)}=\theta^*$ and the chain jumps to the new state. Otherwise, $\theta_{(t+1)}=\theta_{(t)}$.
 
The choice of the transition kernel $Q(\theta^*|\theta_{(t)})$ is crucial to guarantee the rapid convergence of the chain. We opt for the popular choice of a multivariate Normal distribution $\mathcal{N}(0,\Sigma)$ centered on the current state and with a covariance matrix $\Sigma$ which determines the size and orientation of the jumps, so that:
\begin{align}
\theta^*=\theta_{(t)}+\zeta_{(t+1)},
\end{align}
where $\zeta_{(t+1)} $ follows $ \mathcal{N}(0,\Sigma)$.
 
\begin{figure*}
  \includegraphics[width=\textwidth]{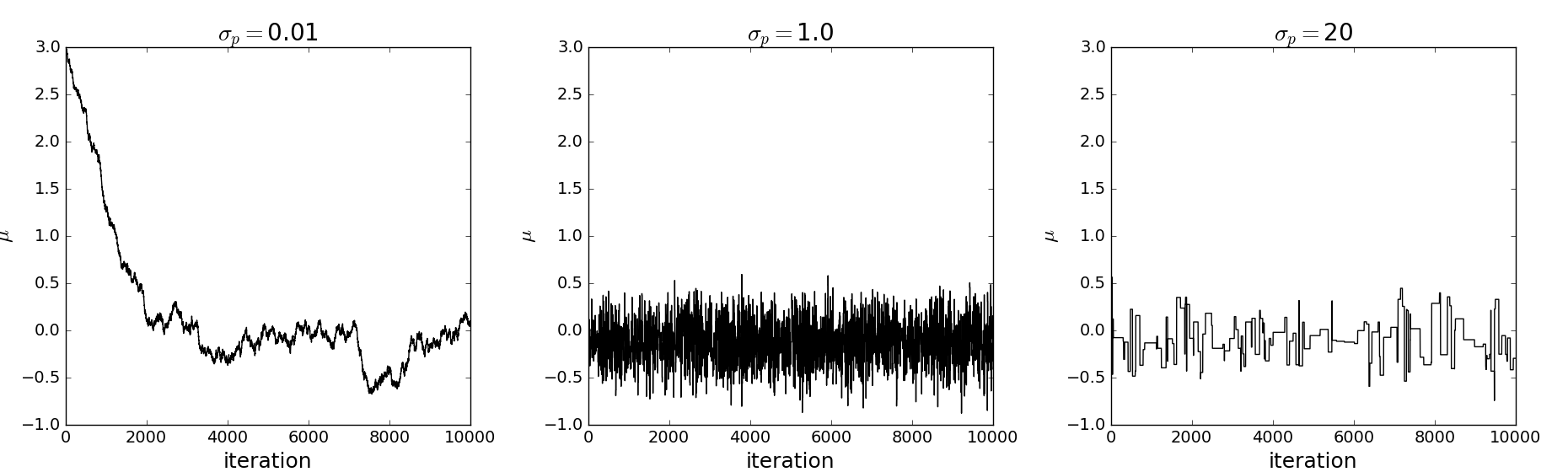}
  \caption{Traceplots depicting three typical situations that can arise in a standard MCMC chain with the (non-adaptive) Metropolis-Hastings algorithm. The input data is a set of 20 points normally distributed with mean 0 and standard deviation 1. The parameter to infer is the mean $\mu$ of the input data distribution. The prior is a Normal distribution with mean 0 and standard deviation 1, and the transition kernel is a Normal distribution centered on the current state and width $\sigma_p$. In each case the chain starts from $\mu=3$ and is run for 10000 iterations. The target distribution sampled is the same, but the width of the proposal distribution, thatis, the jump size, is different for each case. Left panel: The jump size is too small. The burn-in phase is very long and a much longer chain is needed to sample the target distribution. Central panel: The jump size is optimal, therefore the target distribution is well sampled. Right panel: The jump size is too big. Hence the chain spends a lot of iterations in the same position, which makes the sampling of the target distribution inefficient.}
  \label{fig:traceplot_std}
\end{figure*}

A good way to assess convergence speed is to monitor the acceptance rate, that is, the fraction of accepted samples over previous iterations. The acceptance rate is mainly influenced by the covariance matrix of the transition kernel $\Sigma$. If the jump sizes are too high, the acceptance rate is too low, and the chain stays still for a large number of iterations. If the jump sizes are too small, the acceptance rate is very high but the chain needs a high number of iterations to move from one region of the parameter space to another. These situations are illustrated in \fg\ref{fig:traceplot_std}. The desired acceptance rate depends on the target distribution, and there is no universal criterion for its optimization, but Roberts et al (1997) proved that for any $d$-dimensional target distributions (with $d \geq 5)$ with independent and identically distributed (i.i.d.) components, optimal performance of the Random Walk Metropolis algorithm is attained for an asymptotic acceptance rate of 0.234.
 
As the modeling process is very time-consuming and the dimensionality of the problem may be high, we cannot afford to rely on trial and error to find the roughly optimal covariance matrix. We therefore opt for an adaptive MCMC scheme to limit user intervention as much as possible and achieve fast convergence. In the AP algorithm proposed by \citet{Haario1999}, the covariance matrix of the Gaussian kernel $\Sigma$ is tuned on-the-fly every fixed number of iterations using previously sampled states of the chain, and it therefore ``learns'' the target distribution covariance matrix. In our custom version of the algorithm, every $S$ iterations the empirical covariance matrix from every different accepted state of the $N_{last}$ iterations is computed. We then add a fixed diagonal matrix with elements very small relative to the empirical covariance matrix elements, set to $10^{-6}$, to prevent it from becoming singular (\citealt{haario2001}) while not impacting the results much (but to which extent remains presently an open question). The choice of $S$, also called the update frequency, is left to the user and weakly influences the performance of the algorithm, so we set it arbitrarily to 500. As for $N_{last}$, we set it to 50 in order to minimize the chance of the covariance matrix being strongly influenced by a potential rapid evolution of the last few states. 
 
In order to be able to converge in any case, a Markov chain must be ergodic. A stochastic process is said to be ergodic if its statistical properties can be retrieved by a finite random sample of the process. It is well known that adaptation can perturb ergodicity (see, e.g., \citealt{Andrieu2005}). In order to ensure that an adaptive sampling algorithm has the right ergodic properties, and hence converges to the right distribution, it must verify the Vanishing Adaption condition: the level of adaption must asymptotically depend less and less on previous states of the chain. \citet{Haario1999} showed that the AP algorithm is not ergodic in most cases. To tackle this issue, \citet{haario2001} later released a revised version of their algorithm: the Adaptive Metropolis (AM) algorithm. In the AM algorithm, instead of using a fixed number of previous states, the proposal distribution covariance matrix is computed using all the previous states, which solves the ergodicity problem of the AP algorithm. However, we show in section \ref{sec:application} that our custom implementation of the AP algorithm still yields robust results to our problem.

 \subsection{Prior}
  \label{sec:prior}
 
In any Bayesian inference problem, the choice of the prior distribution $P(\theta)$ is of crucial importance, because different prior choices can result in different posterior distributions from the same data. Without any information on what parameter values most probably explain our data, our choice by default is that of an uninformative prior, that is, a multivariate continuous uniform distribution whose boundaries are chosen according to the limits currently given for each parameter in the literature. The uniform prior is defined as:
 
\begin{align}
P(\theta)= \begin{cases}
    \prod\limits_{i=1}^{N_p} \frac{1}{d_{i}-c_{i}}    & \text{  if }d_i \leq \theta_i \leq c_i\text{ } \forall i \in [1,N_p]\\
    0    & \text{ else},
    \end{cases}
\end{align} 
where $c_i$ and $d_i$ are the lower and upper limit of the PDF for parameter $i$ and $\theta=(\theta_1,\theta_2,...,\theta_{N_p})$ is the parameter values vector. \\
 
If more precise information is available on a given subset of parameters, a convolution with a more informative PDF (e.g., Normal, Beta...) can be performed, but in any case a finite interval is needed in order to provide the source generation software with realistic input parameters. In fact, an infinite interval can result in situations in which no galaxies are generated by the model, or conversely when too many galaxies are generated, which would dramatically increase the computing time. 
 
\subsection{Acceptance probability}
\label{sec:acceptance}

In practice, one uses the ratio of the posterior density at the proposed and current states to measure the acceptance probability. More specifically, we use the difference between the log of these quantities in order to avoid floating-point numbers precision problems when dealing with very small probabilities. In log probability space, Bayes' theorem (cf \eq \ref{eq:bayes}) becomes:
 \begin{align}
\label{eq:bayes2}
\ln P(\theta|\mathcal{D}) \propto \ln P(\theta) + \ln P(\mathcal{D}|\theta),
\end{align}
where $\mathcal{D}$ is the input data, $P(\theta|\mathcal{D})$ is the posterior, $P(\theta)$ is the prior defined in section \ref{sec:prior}, and $P(\mathcal{D}|\theta)$ is the auxiliary likelihood defined in \eq \ref{lnL}.
 
The target distribution can have a complex shape and if no particular precaution is taken, our sampling algorithm is not immune to getting stuck in a local maximum of likelihood. To tackle this issue, \citet{10.2307/1690046} exploited the analogy between the way a heated metal cools and the search for a global optimum of a function. In the so-called simulated annealing algorithm, the acceptance probability $a$ depends on a ``temperature'' parameter $\tau$, initialized at high value and slowly decreasing over the iterations. In this scheme, the higher the temperature, the higher the algorithm is prone to accept large moves and to get away from a nearby local maximum:
\begin{align}
a= \begin{cases}
    \exp -\frac{\ln P(\theta_{(t)}|\mathcal{D})-\ln P(\theta^*|\mathcal{D})}{\tau}   & \text{  if } \ln P(\theta^*|\mathcal{D})-\ln P(\theta_{(t)}|\mathcal{D}) < 0 \\
    1    & \text{ if } lnP(\theta^*|\mathcal{D})-\ln P(\theta_{(t)}|\mathcal{D}) \geq 0\\
    \end{cases}
\label{simannealaccept}
\end{align}
where $\ln P(\theta_{(t)}|\mathcal{D})$ and $\ln P(\theta^*|\mathcal{D})$ are respectively the log of the posterior density at the current (\ie, at iteration t) and the proposed state. In other words, if a proposition is considered more probable, it is accepted. Otherwise, it is accepted with probability $a$ (defined in \eq \ref{simannealaccept}). To perform the latter operation in practice, a uniformly distributed random number $R_N$ is drawn in the interval $[0,1]$. If $R_N<a$, the jump is accepted. As expected, for $\tau=1$, the acceptance probability is the same as that of the Metropolis-Hastings algorithm in \eq \ref{accratemh} for the particular case of a symmetric proposal distribution, that is, when $Q(\theta_{(t)}|\theta^*)=Q(\theta^*|\theta_{(t)})$ .
 
Because of the intrinsic stochasticity of our model, many realizations of the model at the same state $\theta_{(t)}$ can lead to many $\ln P(\theta_{(t)}|\mathcal{D})$ values. Therefore, artificial local maxima of the target distribution appear, because each iteration relies on a single realization of the model. The simulated annealing algorithm was designed to find the global maximum of the target distribution without knowing the posteriori distribution, and this requires us to lower $\tau$ in a user-defined scheme. But our goal is distinct as we need to freely explore the parameter space landscape in order to estimate the full posterior distribution. The main constraint for $\tau$ is to be comparable to the posterior density difference resulting from the jump. Here we define it as the root mean square (RMS) of the current state, as suggested by \citet{Mehrotra1997}. In that scheme, a high noise level or a small difference between the proposed and the current state leads to a higher probability of jumping to this state. 
 
\begin{figure}
  \includegraphics[width=0.49\textwidth]{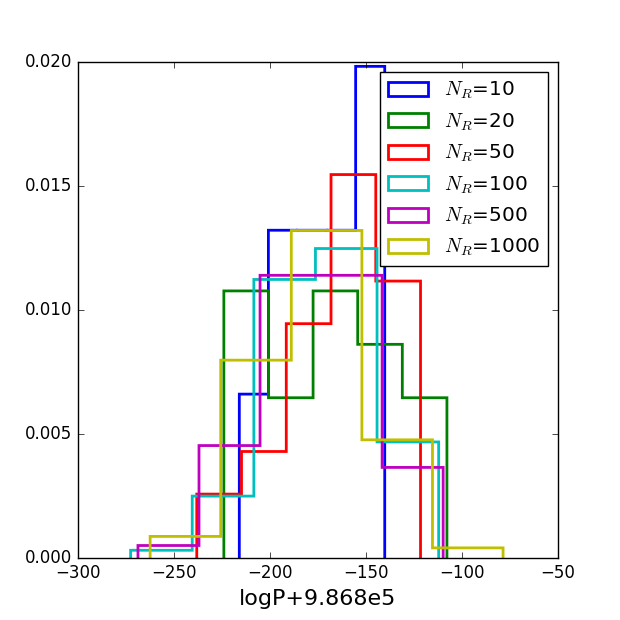}
  \caption{Normed distribution of $\ln P$ for various numbers of realizations $N_R$ of the model. Each distribution is generated in the conditions of the ``Fattening E'' case, at ``true'' input values (cf Table~\ref{tab:popparams}) and with the same seed for galaxy generation in {\sc Stuff}. Standard deviation of the distributions do not appear to differ significantly. We conclude that 20 realizations of the model are enough to characterize the order of magnitude of RMS.}
  \label{fig:noisemag}
\end{figure}
 
\begin{figure}
  \includegraphics[width=0.49\textwidth]{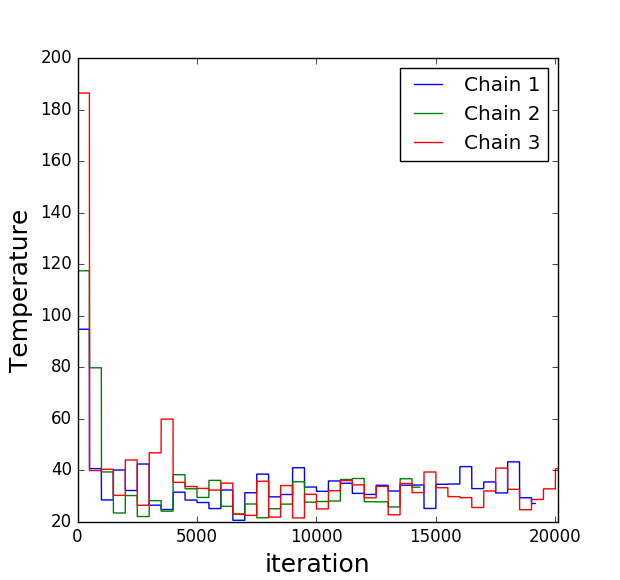}
  \caption{Temperature evolution with the number of iterations of the MCMC process in the ``Fattening E'' case described in \sct\ref{subsec:1popE}. Here the temperature is computed every 500 iterations at current state with 20 realizations of the model. We note that for each chain, the temperature values quickly converge to the level of noise of the model near input values.}
  \label{fig:Tevolrun120}
\end{figure}

The temperature is computed every $S$ iterations by running an empirically-defined number of realizations $N_R$ of the model at the current state $\theta_{(t)}$, storing every $\ln P(\theta_{(t)}|\mathcal{D})$ value returned in a vector, and computing the standard deviation of the resulting distribution. In the application below, we find that 20 realizations are sufficient to give a reasonable estimate of the RMS (cf \fg\ref{fig:noisemag}) and that the temperature quickly reaches a stationary distribution at a relatively low level $\tau\simeq30$, after the first few $10^3$ iterations (cf \fg\ref{fig:Tevolrun120}).
 
\subsection{Initialization of the chain}
 
The initial state $\theta_{(0)}$ is drawn randomly from the prior distribution (see \sct\ref{sec:prior}). The initial position will only affect the speed of convergence, because the final distribution shall not depend on the initial position, if the chain converges. The initial temperature is then computed from this state. As for the proposal distribution, it is initialized so that no direction in the parameter space is preferred by the sampling algorithm at first. The initial covariance matrix is therefore diagonal, whose non-zero elements are set to:
 
\begin{align}
\label{eq:initCii}
C_{ii}=\frac{u_i-l_i}{E} \text{ } \forall i  \in [1,N_p],
\end{align}
where $u_i$ and $l_i$ are respectively the upper and lower bounds of the prior distribution for parameter $i$, $N_p$ the number of parameters, and $E$ a value set empirically to 200 in order to ensure reasonable acceptance rates at the beginning of the chain. According to \citet{Haario1999}, the adaptive nature of the algorithm implies that the choice of $E$ should not influence the output of the chain.
 
\section{Convergence diagnostics}
\label{sec:convergence}
 
The goal of an MCMC chain is to reach a stationary distribution that is supposed to be representative of the target distribution. Unfortunately there is no theoretical criterion for convergence: in other words it is impossible from a finite MCMC chain to assess convergence with certainty. Many convergence diagnostics have been developed (the reader can find an extensive review of those and a comparison of their relative performances in, e.g., \citealt{Cowles1996}), but these diagnostics can only tell if a chain has not converged. So in order to have confidence in the convergence of the chains, we must perform multiple diagnostics. 
 
The first check is carried out by visual inspection of the trace plot for each parameter. Trace plots are used to diagnose poor mixing, that is, when the chain is highly autocorrelated, or slow sampling caused by too small a step size, which suggests that the majority of the MCMC output is not representative of the target distribution (see \fg\ref{fig:traceplot_std}). We also use trace plots to estimate the length of the burn-in phase. The latter is determined by eye, by a rough estimate of the minimum number of iterations $D$ necessary for all the parameters to reach a seemingly stationary distribution. We then discard the $D$ first iterations, where $D$ depends on the chain.
 
Finally, one of the most popular convergence diagnostics is a test proposed by \citet{Gelman1992}. Given $m$ chains $\{ \theta_{(t)}^j \}$ ($j=1,...,m$ and $m \geq 3$, and typically $\sim$ 10), each of length $n$ after discarding  burn-in ($t=1,...,n$) and with different starting points, the test compares the variance between the mean values of the m chains B and the mean of the m within-chain variances W:
 
\begin{align}
B = \frac{n}{m-1} \sum_{j=1}^m (\bar{\theta}_.^{j} - \bar{\theta}_{..})^2,
\end{align}
 
\begin{align}
W = \frac{1}{m} \sum_{j=1}^m \left[ \frac{1}{n-1} \sum_{i=1}^n (\theta_{(i)}^{j} - \bar{\theta}_.^{j})^2 \right],
\end{align}
where $ \bar{\theta}_.^{j} =\frac{1}{n} \sum_{t=1}^n \theta_{(t)}^j
$ is the mean value of chain j, and 
$ \bar{\theta}_{..} = \frac{1}{m} \sum_{j=1}^m \bar{\theta}_.^{j}
$ is the average value over the m chains.

An overestimate of the true marginal posterior variance is given by the unbiased estimator
 
\begin{align}
\hat{V} = \frac{n-1}{n} W + \frac{1+m}{nm} B.
\end{align}
 
Finally convergence is estimated using the potential scale reduction factor (PSRF) $\hat{R}$:
 
\begin{align}
\hat{R} = \frac{\hat{V}}{W}.
\end{align}
 
Here we use the Gelman Rubin diagnostic implemented in this form in the PyMC package (\citealt{Patil2010}) to perform our convergence tests, and we consider that convergence has been reached if $\sqrt{\hat{R}} < 1.1$ for all model parameters (\citealt{brooks1998some}); otherwise, more iterations are performed until the criterion is met.

 
\section{Application to a toy model}
\label{sec:application}
 
As a proof-of-concept of the method, we apply our pipeline to a selection of idealized cases, where the ``observed'' data is a synthetic image containing one or two populations of galaxies generated by a set of known input parameters of the {\sc Stuff} model. Our goal is to infer the values of the input parameters in this framework. 
 
\subsection{Simulated survey characteristics}
 
\begin{table*}[ht]
\centering
\caption{\label{tab:characteristics} Imaging characteristics of the CFHTLS+WIRDS surveys \\ used for {\sc SkyMaker}}
\begin{tabular}{l c c c}
  \hline
  \hline
  Passband & \textit{u} & \textit{i} & \textit{$K_s$}  \\
    \hline
  Image size [pixels] & $19354\times 19354$ & $19354\times 19354$ & $19354\times 19354$ \\
  Effective gain [$e^-$/ADU] & 74590 & 6807 & 2134  \\
  Well capacity [$e^-$] & $\infty$ & $\infty$ & $\infty$  \\
  Saturation level [ADU] & 6465 & 4230 & 110884 \\
  Effective read-out noise [$e^-$] & 4.2 & 4.2 & 30 \\
  Total exposure time [s] & 1 & 1 & 1 \\
  Zero-point magnitude [``ADU/s''] & 30 & 30 & 30 \\
  Effective wavelength [$\mu$m] & 0.381 & 0.769 & 2.146 \\
  Sky level [AB mag/arcsec$^2$] & 22.2 & 20.0 & 15.4 \\
  Seeing FWHM [arcsec] & 0.87 & 0.76 & 0.73 \\
  \hline
\end{tabular}
 
\end{table*}
 
As data image, we choose to reproduce a full-sized stack of the CFHTLS Deep field \citep[e.g.,][]{Cuillandre2006}. The CFHTLS Wide and Deep fields offer carefully calibrated stacks with excellent image quality. Covering 155 deg$^2$ on the sky in total, the Wide field allows for a detailed study of the large scale distribution of galaxies. As for the Deep field, which covers 4 deg$^2$ in total, it beneficits from long time exposures (33 to 132 hours), which ensure reliable statistical samples of different populations of bright galaxies up to z $\sim$ 1. Each stack of the CFHTLS Deep field is a 19,354 $\times$ 19,354 pixel image covering 1 deg$^2$ on the sky. We simulate one stack of the Deep field in three bands: Megacam \textit{u} and \textit{i} from the CFHTLS, and the WIRcam $K_s$ infrared channel from the WIRcam Deep Survey (WIRDS) that covers part of the CFHTLS Deep fields. In accordance with CFHTLS product conventions, the image exposure time is normalized to one second and the AB magnitude zero-point is 30. The overall characteristics of the simulated images are summarized in Table~\ref{tab:characteristics}. 

The {\sc SkyMaker} PSF model for the CFHTLS image is generated within the software. The \textit{aureole} simulation step is deactivated to speed up the image generation process. For the same reason, we exclude from the {\sc Stuff} list all galaxies with apparent magnitudes in the reference band brighter than 19 or fainter than 30, in order to avoid simulating both very large and very numerous galaxies. There is no stellar contamination, as {\sc Stuff} does not yet offer the possibility to simulate realistic star fields. 
 
\subsection{Source extraction configuration}
 
{\sc SExtractor} is configured according to the prescription of the T0007 CFHTLS release documentation \citep{manT0007}. We use it in double image mode, with the \textit{i}-band image as the detection image, and the background is estimated and subtracted automatically with a 256$\times$256-pixels background mesh size. In order to optimize the detectability of faint extended sources, detection is performed on the images convolved with a 7$\times$7 pixels Gaussian mask having a full width at half maximum (FWHM) of three pixels, that approximates the size of the PSF and acts as a matched filter. Finally, the detection threshold is set to $1.2$ times the (unfiltered) RMS background noise above the local sky level.
 
In order for the results concerning faint sources near the detection limit not to depend too closely on the details of noise statistics, all negative fluxes and radii are clipped to 0 after extraction.
 
\subsection{Pipeline configuration}
 
\begin{table*}
\centering
\caption{\label{tab:priors} Uniform prior boundaries for the parameters of the luminosity and size functions, and their evolution with redshift.}
\begin{tabular}{c c c c c c c c c c}
  \hline
  \hline
  Parameter & $\phi^*$ & $M^*$ & $\alpha$ & $\phi_e$ & $M_e$ & $M_{knee}$ & $r_{knee}$ & $\gamma_b$ \\
  \hline
  lower bound & $10^{-7}$ & -22 & -2.5 & -3 & -2.5 & -21 & 0 & -2 \\
  upper bound & $10^{-2}$ & -17 & 0 & 2 & 0 & -19 & 3 & 0 \\
  \hline
\end{tabular}
\tablefoot{All the parameters above are given for $H_0=100~\mathrm{km}.\mathrm{s}^{-1}.\mathrm{Mpc}^{-3}$. \\}
\end{table*}
 
We adopt non-informative, uniform priors for the free parameters of all the considered models, with boundaries defined in Table~\ref{tab:priors}. The boundaries are chosen to prevent the pipeline from exploring non physical domains, such as a very steep LF faint end, which leads to an unreasonably high number of generated galaxies and dramatically increases the computing time. We select the least constraining prior possible, which corresponds to a large interval around generally accepted values, such as the values reviewed in \citet{DeLapparent2003} for example.
 
\begin{table}
\caption{Parameters of the dynamic range reduction function used in \eq\ref{eq:reducrange}.}
\centering
\begin{tabular}{c c c c}
\hline
\hline
Filter & \textit{u} & \textit{i} & $K_s$ \\
\hline
$\sigma_{FLUX\_AUTO}$ & 3.4 & 3.6 & 54.0 \\
$\sigma_{FLUX\_RADIUS}$ & 3.5 & 2.7 & 2.6 \\
$\kappa_{c}$ & 10 & 10 & 10 \\
\hline
\end{tabular}
 
\label{tab:dynrange}
\end{table}
 
To perform the dynamic range compression as defined in \sct\ref{subsec:reddynrange}, we need an estimate of the noise level in the conditions of a CFHTLS Deep field. To that end, we use the population of $\sim 10^4$ pure bulge elliptical galaxies described in \sct\ref{subsec:1popE} and apply the recipe described in \sct\ref{subsec:reddynrange}. The resulting parameters for the dynamic range reduction function in the \textit{ui$K_s$} filters are summarized in Table~\ref{tab:dynrange}. For the various cases considered in this article, we use for all galaxy populations the $\sigma_{FLUX\_AUTO}$ and $\sigma_{FLUX\_RADIUS}$ values measured for the elliptical galaxies. \\

 
We consider two cases in the following sections: the first contains two types of galaxies, a mix between ellipticals and lenticulars, and late-type spirals, which undergo both luminosity and size evolution. But we limit the inference to the LF shape and evolution parameters for both populations. The second case focuses on a single population of pure bulge ellipticals, but this time the inference is performed on both the LF and the distribution of effective radii (both including the evolution parameters).
 
\subsection{Multi-type configuration: luminosity evolution}
\label{subsec:2_pop}
 
\begin{table*}
\centering
\caption{\label{tab:popparams} Characteristics of the galaxy test populations.}
\begin{tabular}{c c c c c c c c c c c c}
  \hline
  \hline
  Population & $SED_{b}$\tablefootmark{a} & $SED_{d}$\tablefootmark{a} & $\phi^*$ [$h^3$ Mpc$^{-3}$] & $M^*$ & $\alpha$ & $\phi_e$ & $M_e$ & $B/T$ & \text{T}\tablefootmark{b} & $\alpha(T)$ & Number\tablefootmark{c} \\
  \hline 
  Multi-type: E/S0 & E & E & 0.003 & -19.97 & -0.5 & -1.53 & -1.77 & 0.65 & -5 & 0.0 & 10447 \\
  Multi-type: Sp & E & Scd & 1.4e-4 & -19.84 & -1.3 & 0.03 & -1.95 & 0.2 & 6 & 1.47 & 28281 \\
  Fattening E & E & E & 0.0035 & -19.97 & -0.5 & -1.53 & -1.77 & 1.0 & -5 & 0.0 & 11353 \\
  \hline
 
\end{tabular}
\tablefoot{ The LF parameters are given for $H_0=100~\mathrm{km}.\mathrm{s}^{-1}.\mathrm{Mpc}^{-3}$. \\ \tablefoottext{a}{The disk and bulge SEDs are \citet{Coleman1980} templates.} \\ \tablefoottext{b}{ \citet{DeVaucouleurs59} revised morphological type.} \\
\tablefoottext{c}{Number of sources generated by one realization of {\sc Stuff}.}}
\end{table*}
 
Astronomical survey images contain multiple galaxy populations. We need to emulate this situation in order to test the behavior of our pipeline in realistic conditions. To do so we use as input data a simulated CFHTLS Deep image in \textit{ui$K_s$} containing two types of galaxies: a population of early-type galaxies (an average between E ans S0) of morphological type T=-5 and a population of late-type spirals (Sp) of morphological type T=6. We rely on published results to define these populations. 
Using data from SDSS, the 2dF Galaxy Redshift Survey, COMBO-17, and DEEP2, \citet{faber07} split their distribution of galaxies into two populations split by color, using the rest-frame $M_B$ versus $U-B$  color-magnitude diagram: a blue population and a red population. We use their derived evolving LF parameters to build an E/S0 and Sp populations. The detailed conversion process from the LF parameters of \citet{faber07} to the values used in {\sc Stuff} (which include a magnitude system conversion, a band transformation, and a cosmological correction) is provided in Appendix \ref{app:convfaberstuff}. This provides us with values for $M^*$ (LF characteristic magnitude) and the evolution parameters $M_e$ and $\phi_e$ for both populations (see Section \label{subsec:stuff}). 
 
The B/T ratios in the g adopted reference band are determined using the distribution of B/T in g-band as a function of  morphological type from EFIGI (Extraction of Idealized Forms of Galaxies in Image processing) data (\citealt{Baillard2011}, de Lapparent, private communication). To limit run time, the $\phi^*$ values for each population are set to have $\sim 4 \times 10^{4}$  galaxies in total generated quickly by {\sc Stuff} in the field area. In this scheme, we have $\sim 10^4$  E/S0, and $\sim 3 \times 10^4$ Sp, which corresponds to a $\phi^*$ value for each population of ten times lower than the values given by \citet{faber07}. We indeed do not match the number counts of a CFHTLS Deep field as it would lead to unreasonable computing time: reproducing realistic number counts over a full Deep field would actually imply {\sc Stuff} generating a number of galaxies one order of magnitude higher for E/S0 and Sp, and also adding a population of $\sim 10^5$ Irr which dominates the number counts fainter than 22 to 24 mag, depending on the filter.
 
The input parameters used to generate both populations are listed in Table~\ref{tab:popparams}. The parameters to infer in this case are the five evolving LF parameters for each of the populations: $\phi^*$, $M^*$, $\alpha$, $\phi_e$, and  $M_e$, that is a total of ten parameters (we do not infer the size distribution and evolution parameters). The observables are the {\sc SExtractor} \textbf{FLUX\_AUTO} in each of the three passbands, which leads to a three-dimensional observable space. Using ten bins for each observable as indicated in \sct\ref{subsec:binning}, we obtain a total number of $10^3$ bins in the observable space. Over the $\sim 4 \times 10^4$ galaxies generated by {\sc Stuff}, we find that $\sim 2 \times 10^4$ are extracted with {\sc SExtractor}. The number of extracted galaxies per bin is presented in \fg\ref{fig:histbins}. 
 
\subsection{Results of the ``multi-type'' configuration}
 
We run the pipeline on a hybrid computing cluster of seven machines totaling 152 central processing unit (CPU) cores. We launched three chains in parallel for 18,357, 18,565, and 16,211 iterations respectively, with randomly distributed starting points, using 50,400 CPU hours in total. The burn-in phase is estimated by visual examination of the trace plot. All the iterations before the upper and lower envelope of the trace becomes constant for all the chains and for all the parameters simultaneously are discarded as burn-in, which in the case under study corresponds to the first $10^4$ iterations. Then convergence over the $f$ last iterations of each chain is assessed based on the Gelman-Rubin test (cf Table~\ref{tab:gelrub}), where $f$ is the minimum length over the three chains after burn-in, as the convergence test requires the same number of iterations for all the chains: $f=6,211$ iterations. Table~\ref{tab:gelrub} lists the results of the Gelman-Rubin test, which suggest that all the chains have converged to the same stationary distribution.

\begin{figure*}
  \includegraphics[width=\textwidth]{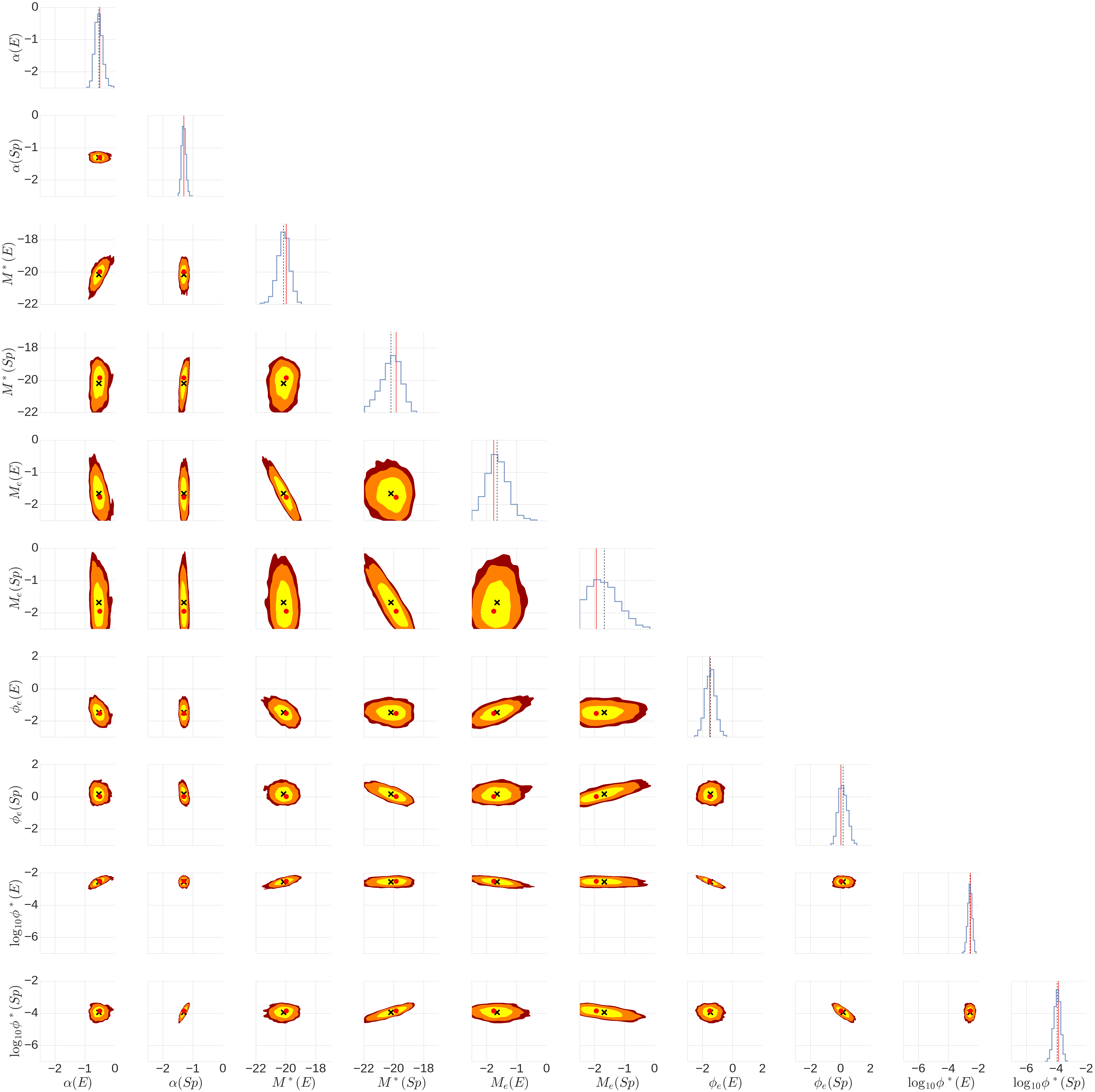}
  \caption{Joint posterior distribution resulting from the ``Multi-type'' test described in \sct\ref{subsec:2_pop}. The diagonal plots show the marginal distribution for each parameter (the projection of the posterior onto that parameter). Each panel is bounded by the prior range values.  The dark red, orange, and yellow areas in the contour plots represent the 99\%, 95\%, and 68\% credible regions respectively. The black crosses and red dots are the mean of the posterior and input true value respectively. In the marginalized posterior panels, the black dotted and red lines represent the posterior mean and the true value respectively.}
  \label{fig:posterior140}
\end{figure*}

The final joint posterior distribution is the result of the combined accepted states of all the chains run after burn-in. The posterior PDF plot is shown in \fg\ref{fig:posterior140}: it contains 3,017 accepted iterations out of 23,132 propositions, corresponding to an overall 13\% acceptance rate after burn-in. 
The graph shows that the ``true'' input values all lie within the $68\%$ credible region, which in Bayesian terms means that there is a 68\% probability that the model value falls within the credible region, given the data. Summary statistics of the posterior PDF are listed in Table~\ref{tab:resultinfer}.
As the pipeline generates constraints that are consistent with the input parameters, we therefore conclude that our approach can be used to perform unbiased inference on the photometric parameters of galaxies using two broad classes of galaxy types given non-informative priors.
 
Moreover, we find in \fg\ref{fig:posterior140} some strong correlations or anti-correlation between various pairs of parameters, that are symptomatic of the degeneracies in the parameters for our specific set of observables (fluxes). For example, a strong anti-correlation is found between $M^*$ and $M_e$ in the two populations. This can be explained by the fact that a brighter (lower) $M^*$ population at $z=0$ can be partly compensated by a shallower (higher) redshift evolution.

 
\begin{table*}
\centering
\caption{\label{tab:sizeevolparams} Size parameters for the bulge and disk of each galaxy test population}
\begin{tabular}{c c c c | c c c}
  \hline
  \hline
   \multicolumn{4}{c}{Disk} & \multicolumn{3}{c}{Bulge} \\
  $\beta_d$ & $r^*_d$ [$h^{-1}$kpc] & $\gamma_d$ & $\sigma_{\lambda}$ & $M_{knee}$ & $r_{knee}$ [$h^{-1}$kpc] & $\gamma_b$ \\
  \hline
  -0.214 & 3.85 & -0.80 & 0.36 & -20.0 & 1.58 & -1.00 \\
  \hline
\end{tabular}
\tablefoot{All the parameters above are given for $H_0=100~\mathrm{km}.\mathrm{s}^{-1}.\mathrm{Mpc}^{-3}$, and ``b'' refers to bulge, and ``d'' to disk. \\}
\end{table*}

\subsection{Fattening ellipticals: size and luminosity evolution}
\label{subsec:1popE}
We then test whether our pipeline can also infer the characteristic size and size evolution of galaxies. Because of memory limitations, we perform this test in a simplified framework. We use as input data a CFHTLS image in \textit{ui$K_s$} containing $\sim 10^4$ E/S0 (pure bulge) galaxies generated with {\sc Stuff}. The input photometric parameters are listed in Table~\ref{tab:popparams} and those for bulge size are listed in Table~\ref{tab:sizeevolparams}. The parameters to infer are the five evolving LF parameters, as well as three parameters governing the bulge distribution and evolution: $M_{knee}$, $r_{knee}$, and $\gamma_b$ (as defined in \sct\ref{subsec:stuff}). That is a total of eight parameters. No extinction is included in this case. As the size evolution parameters cannot be retrieved with the photometric information only (\textbf{FLUX\_AUTO}), the \textbf{FLUX\_RADIUS} parameters of {\sc SExtractor} for all galaxies in each passband are added to the observables space. This leads to a six-dimensional observable space. Over the $\sim 10^4$ E generated by {\sc Stuff}, we find that $\sim 7 \times 10^3$ are found by {\sc SExtractor}. With ten bins as indicated in \sct\ref{subsec:binning}, this results in a total number of bins of $10^6$. The number of extracted galaxies per bin is presented in \fg\ref{fig:histbins}.
 
\subsection{Results of the ``fattening ellipticals'' configuration}
 
\begin{table*}
\centering
\caption{\label{tab:gelrub} Results of the Gelman-Rubin test.}
\begin{tabular}{c c c c c c c c c}
\hline
\hline
Population  & $\log_{10}(\phi^*)$ & $M^*$ & $\alpha$ & $\phi_e$ & $M_e$ & $M_{knee}$ & $r_{knee}$ [$h^{-1}$kpc] &  $\gamma_b$ \\
\hline
Multi-type: E/S0  & 1.015 & 1.006 & 1.014 & 1.012 & 1.005 & $\varnothing$ & $\varnothing$ & $\varnothing$ \\
Multi-type: Sp  &  1.020 & 1.013 & 1.028 & 1.007 & 1.012 & $\varnothing$ & $\varnothing$ & $\varnothing$ \\
Fattening E  & 1.013 & 1.003 & 1.020 & 1.003 & 1.001 & 1.008 & 1.010 & 1.008 \\
\hline
\end{tabular}
\tablefoot{The values of $\sqrt[]{R}$ are obtained using 3 chains for each case, whose burn-in phase for each chain is determined by eye. All values are $< 1.1$, which is a hint that in each case, all the chains have converged to the same distribution. The parameters above are given for $H_0=100~\mathrm{km}.\mathrm{s}^{-1}.\mathrm{Mpc}^{-3}$.}
\end{table*}
 
\begin{figure*}
  \includegraphics[width=\textwidth]{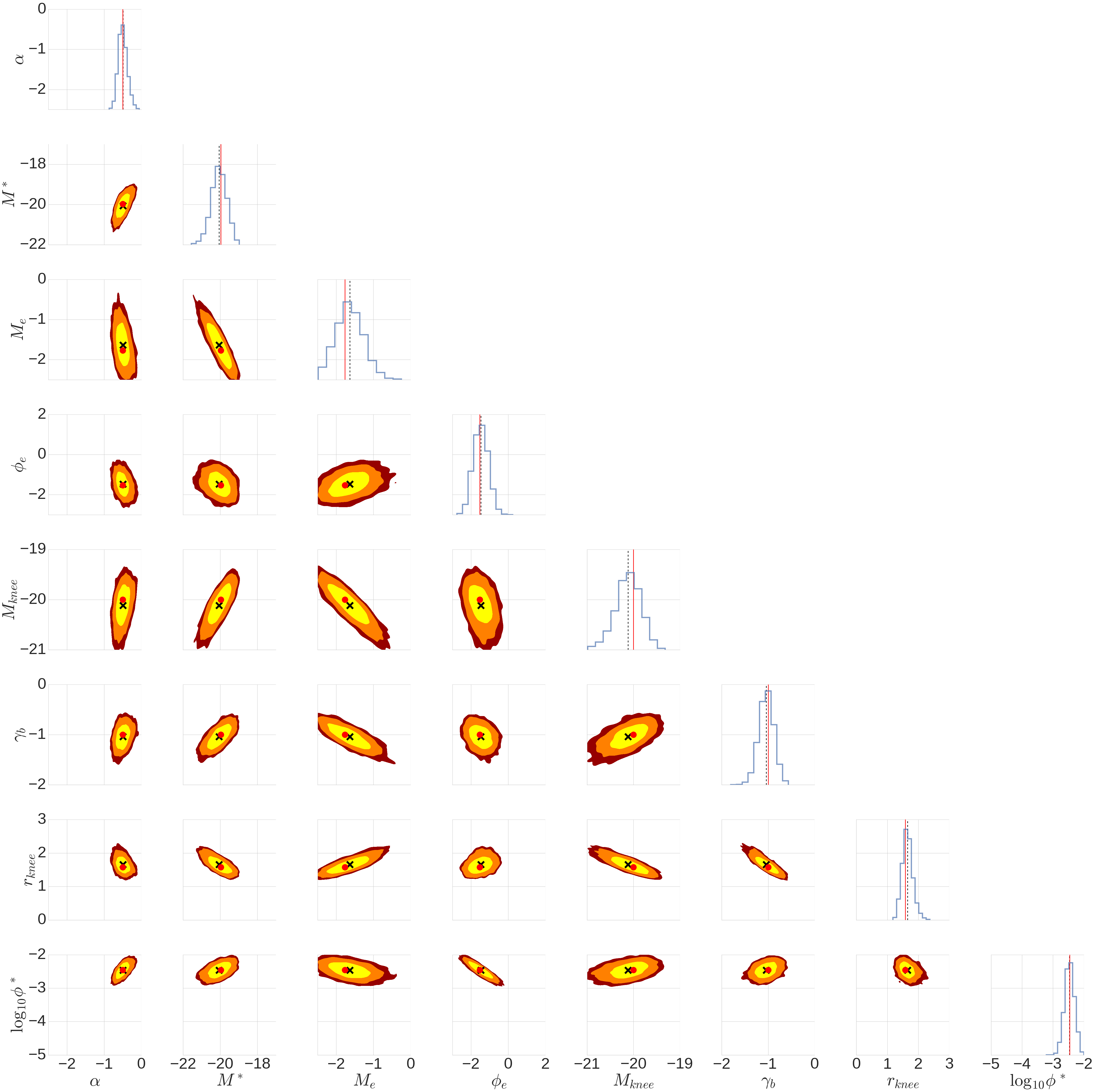}
  \caption{Joint posterior distribution resulting from the ``Fattening E'' test described in \sct\ref{subsec:1popE}. The diagonal plots show the marginal distribution for each parameter. Each panel is bounded by the prior range values.  The dark red, orange, and yellow areas in the contour plots represent the 99\%, 95\%, and 68\% credible regions respectively. The black crosses and red dots are the mean of the posterior and input true value respectively. In the marginalized posterior panels, the black dotted and red lines represent the posterior mean and the true value respectively.}
  \label{fig:posterior120}
\end{figure*}

We run our pipeline with three chains in parallel for 18,898, 14,056, and 20,110 iterations respectively, with uniformly distributed starting points, using 19,656 CPU hours in total. The first $10^4$ iterations of each chain are discarded as burn-in. Convergence is reached over the $f=4,323$ last iterations of each chain, as assessed by the Gelman-Rubin test results displayed in Table~\ref{tab:gelrub}. The resulting posterior distribution is shown in \fg\ref{fig:posterior120}. It contains $6,287$ accepted iterations over $38,064$, which leads to an acceptance rate of 16.5\%. 
 
Each marginalized posterior plot exhibits a main mode, with the peak and the mean almost indistinguishable from the input values. The joint posterior distribution shows that the input values all fall within the $68\%$ credible region. Summary statistics of the posterior PDF are listed in Table~\ref{tab:resultinfer}. Here again, our pipeline produces constraints that are consistent with the true parameters. So we conclude that our pBIL method can reliably infer the luminosity and size distribution of one population of galaxies without any systematic bias.
 
The joint posterior PDF also reveals covariances between parameters.  For instance, the $\phi^*$ and $\phi_e$ parameters are naturally anti-correlated because an increase of $\phi^*$ (at $z=0$) can partially be compensated by a steeper decrease of the normalization with redshift, hence a smaller value of $\phi_e$.
 
\begin{table*}
\centering
\caption{\label{tab:resultinfer} Summary statistics on the marginalized posterior distributions for the galaxy test populations and comparison with the input values. }
\begin{tabular}{c c c c c c c c c c c c}
  \hline
  \hline
 Population & Parameters & Input value & Mean & MAP \tablefootmark{a} & $68\%$ interval & $95\%$ interval & $99\%$ interval\\
   \hline
    & $\log_{10}(\phi^*)$ & -2.52 & -2.56 & $-2.57$ & [-2.72,-2.40] & [-2.85,-2.29] & [-2.95,-2.19] \\
   & $M^*$ & -19.97 & -20.15 & $-20.12$ & [-20.54,-19.74] & [-20.95,-19.38] & [-21.20,-19.08] \\
   Multi-type: E/S0 & $\alpha$ & -0.5 & -0.53 & $-0.54$ & [-0.68,-0.43] & [-0.77,-0.24] & [-0.82,-0.06] \\
   & $\phi_e$ & -1.53 & -1.48 & $-1.37$ & [-1.82,-1.15] & [-2.16,-0.81] & [-2.35,-0.49] \\
   & $M_e$ & -1.77 & -1.66 & $-1.70$ & [-1.99,-1.29] & [-2.36,-0.96] & [-2.46,-0.75] \\
   \hline 
   & $\log_{10}(\phi^*)$ & -3.85 & -3.93 & $-3.96$ & [-4.15,-3.67] & [-4.35,-3.54] & [-4.53,-3.40] \\
   & $M^*$ & -19.84 & -20.18 & $-19.81$ & [-20.68,-19.43] & [-21.59,-19.01] & [-21.93,-18.84] \\
   Multi-type: Sp  & $\alpha$ & -1.3 & -1.30 & $-1.33$ & [-1.37,-1.25] & [-1.42,-1.17] & [-1.44,-1.13] \\
   & $\phi_e$ & 0.03 & 0.18 & $0.15$ & [-0.16,0.47] & [-0.34,0.75] & [-0.48,0.94] \\
   & $M_e$ & -1.95 & -1.68 & $-1.81$ & [-2.39,-1.29] & [-2.49,-0.87] & [-2.49,-0.19] \\
   \hline 
    & $\log_{10}(\phi^*)$ & -2.46 & -2.46 & $-2.51$ & [-2.60,-2.31] & [-2.75,-2.18] & [-2.84,-2.08] \\
   & $M^*$ & -19.97 & -20.04 & $-20.08$ & [-20.41,-19.63] & [-20.84,-19.23] & [-21.15,-19.12] \\
   & $\alpha$ & -0.5 & -0.49 & $-0.51$ & [-0.62,-0.40] & [-0.72,-0.25] & [-0.78,-0.13] \\
   & $\phi_e$ & -1.53 & -1.49 & $-1.56$ & [-1.84,-1.09] & [-2.22,-0.74] & [-2.41,-0.49] \\
   Fattening E & $M_e$ & -1.77 & -1.65 & $-1.61$ & [-2.04,-1.29] & [-2.35,-0.96] & [-2.47,-0.73] \\
   & $M_{knee}$ & -20.00 & -20.10 & $-19.99$ & [-20.32,-19.79] & [-20.68,-19.58] & [-20.90,-19.46] \\
   & $r_{knee}$ & 1.58 & 1.64 & $1.56$ & [1.47,1.77] & [1.34,1.97] & [1.27,2.09] \\
   & $\gamma_b$ & -1.00 & -1.03 & $-1.07$ & [-1.18,-0.87] & [-1.31,-0.71] & [-1.50,-0.65] \\
  \hline
\end{tabular}
\tablefoot{ \tablefoottext{a}{Maximum A Posteriori} \\ The LF parameters are given for $H_0=100~\mathrm{km}.\mathrm{s}^{-1}.\mathrm{Mpc}^{-3}$. }
\end{table*}
 
\section{Comparison with SED fitting}
\label{sec:photoz}
 
\begin{figure}[htb]
  \includegraphics[width=0.49\textwidth]{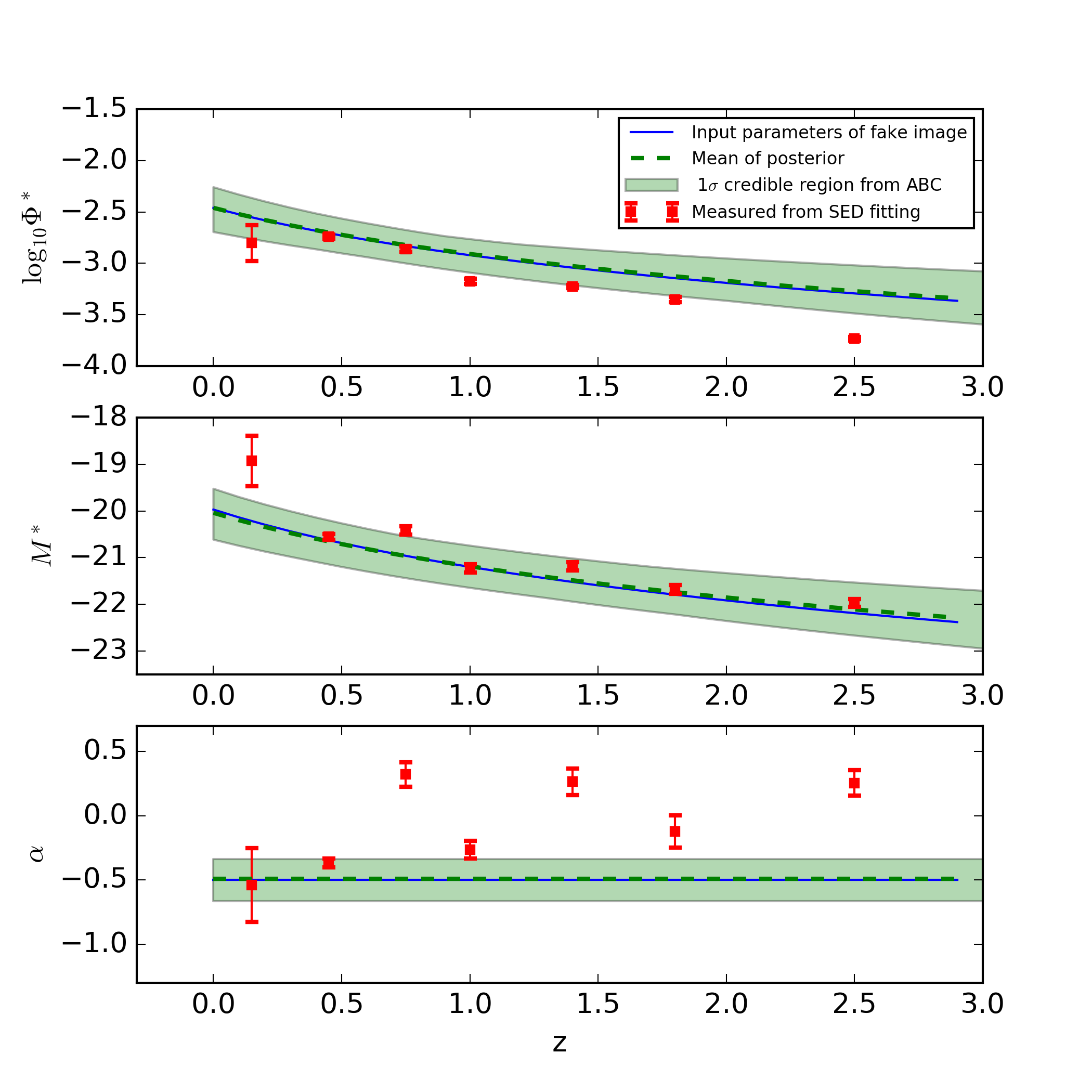}
  \caption{Evolution of the LF parameters as defined in the mock data image (blue solid line) and inferred from the pBIL method in the ``Fattening E'' population described in \sct\ref{subsec:1popE} (the green dotted line is the mean of the posterior and the shaded area represents the $68\%$ credible region), compared to the direct measurement of the LF obtained per redshift bin and estimated using a $V_{max}$ weighting, after determination of the photometric redshifts from SED fitting (red dots).}
  \label{fig:LFevol_and_zphots}
\end{figure}
  As demonstrated above, our pBIL method is efficient at recovering the input parameters used to define the luminosity and size evolutions in the mock CFHTLS image. One may wonder how it compares with the classical, less CPU-expensive method for measuring LFs -- SED fitting-- which provides, from a multi-band photometric catalog, estimates of the photometric redshifts as well as rest-frame luminosities. Luminosity functions can then be derived using independent redshift bins. 

  The simulated field used for this comparison is the ``Fattening E'' sample, with a single population of pure bulges with the \citet{Coleman1980} ``E'' template. The Z-PEG code \citep{LeBorgne2004} is  applied to the $u$, $i$, and $K_s$ photometric catalog obtained with {\sc SExtractor} in the same configuration as described for the pBIL method in \sct\ref{sec:application}. Photometric redshifts are measured together with g-band luminosities for every i-band detected object down to $u_{AB}=30$. The fits were performed using the whole range of SED templates from \citet{Coleman1980}, from E to Irr galaxies. The discrete LFs obtained in each redshift bin were volume weighted with a $V_{max}$ correction at the faint magnitude bins, and a \citet{Schechter1976} function was fitted to the data independently in each redshift bin with a Levenberg-Marquardt algorithm with $\Phi^*$, $M^*$, and $\alpha$ as free parameters.
 
Comparison of the evolution with redshift of the LF parameters between the pBIL approach (green dashed line for mean of posterior, and $68\%$ light green shaded region) and the results from SED fitting (red symbols with error bars) are shown in \fg\ref{fig:LFevol_and_zphots}. As expected, they both roughly follow the trends set by the evolution of the input parameters (blue solid line), with some offsets that can be explained by the fact that SED fitting is done on only three photometric bands. This is clearly a major limiting factor, albeit partly compensated by the choice of the SED templates: the input SED and the templates share a common SED (the ``E'' SED, even if all SEDs from \citet{Coleman1980} are also used for the SED fitting). We believe that this choice is fair because in the pBIL method, the same set of SEDs was also used for data generation and for the inference of LF parameters. 

The significant systematic offset in $\alpha$ from SED fitting compared to the input and pBIL curves in \fg~\ref{fig:LFevol_and_zphots} shows that the faint-end slope parameters $\alpha$ is poorly estimated, with a significant systematic offset at $z\ge0.7$. This is caused by the negative input slope (see Table~\ref{tab:popparams}), which yields few faint galaxies in the sample. Moreover, because of numerous catastrophic outliers in the photometric redshifts (caused by the $u$,$i$, $K_s$-only photometric catalog), there is a mismatch between the true redshift of many faint objects and the redshift bin to which they are assigned. This leads to an underestimate of the error bars on the individual points.
 
For this comparison of the LF parameters between the two approaches, we had to derive the envelop of the LF parameters as a function of redshift for the trace elements of the MCMC chains within the $68\%$ credible region of the parameters space. Of course, the area appears in \fg\ref{fig:LFevol_and_zphots} as much smoother than the individual points derived from SED fitting because the chosen LF model for the inference evolves smoothly with redshift. Still, it is remarkable that the region is tight and almost centered on the values of the true parameters at all redshifts. This is because the various covariances between the five LF parameters of the model tend to narrow down the shaded areas in these graphs, therefore implying that galaxies at all redshifts in the images contribute to constrain the parameters of the model in the pBIL approach.

\section{Discussion}
\label{sec:discussion}
 
One issue of concern in the posterior distributions that we derived with our pipeline (\fgs\ref{fig:posterior140} and \ref{fig:posterior120}) is illustrated by the fact that in \fg~\ref{fig:LFevol_and_zphots}, the $68\%$ shaded region is large compared to the distance between the input parameters (blue solid line) and the mean of the posterior (green dashed line), which are almost indistinguishable in all three graphs. We suspect that this results from an enlargement of the posterior because of the ``temperature'' term that we use in order to circumvent the stochastic nature of each model realization (see \sct\ref{sec:acceptance}). In essence, the model's stochasticity itself (galaxies are randomly drawn from the distribution functions) inevitably contributes to the uncertainty in the posterior. We have no quantitative estimate of this enlargement and we suspect it might be a limiting factor on the precision of the parameter inference. Estimating this enlargement from simulated data would have required us to generate a very large number of realizations for each step of the chain (hence we could have turned off the ``temperature'' term). This would, however, be prohibitive in computing time, even in the considered simple tests performed in this article. We note that using surveys with large statistics in the number of characteristic population of galaxies is, as always, preferable, and should limit this bias.

Moreover, there is room for several technical improvements of our pipeline, in order to guarantee a faster convergence and a more accurate inference:
 
\begin{itemize}
\item As implemented in the present article, our method faces the inevitable curse of dimensionality. In fact, as we bin each observable over ten intervals, for each observable added the hyper-dimensional number of bin increases by one order of magnitude. This limits our approach to a restricted number of observables in order to prevent memory errors. In order to adapt this method to higher numbers of observables, we may have to change our strategy and bin projections of the datasets instead of binning the complete observable space, with the drawback of losing mutual information. 
\item Instead of binning the distribution of observables, whose results depend on the bin edges and bin width, a more reliable method for density estimation for multivariate data is Kernel Density Estimation (KDE). KDE transforms the data points into a smooth density field, and alleviates the dependence of the results on the bin edges by centering a unimodal function with mean 0, the kernel, on each data point. In practice, KDE is more computationally expensive than binning, and also requires some level of hand tuning, in the form of the right kernel function and the optimal bandwidth, which in KDE is the analog of bin width in histograms.
\item  The mean runtime of an MCMC chain in the context of the test cases described in \sct\ref{sec:application} is approximately two weeks. Up to 50\% of this runtime is currently spent in job scheduling latencies for each iteration  (as shown in \fg\ref{fig:bench}). A more integrated approach, based, for example, on Message Passing Interface (MPI) and operating only in memory might reduce those latencies. The next step would be to increase computational efficiency by offloading the most time-consuming image rendering and source extraction tasks to graphics processing units (GPUs), especially convolutions and rasterizations.
\item We emphasize that on the order of $10^4$ iterations is needed to attain convergence in the test cases studied. Considering the high computational cost of this approach, one may wonder how to attain faster convergence in realistic frameworks. In that regard, \citet{gutman2016}, who explored the computational difficulties arising from likelihood-free inference methods, proposed a strategy based on probabilistic modeling of the relation between the input parameters and the difference between observed and synthetic data. This approach would theoretically reduce the number of iterations needed to perform the inference.
\end{itemize}
 
Finally, more realistic mock astrophysical images are required before running our pipeline on real survey data:
 
\begin{itemize}
\item The addition of a likely stellar field to the simulated images would contaminate the source extraction process in a realistic way. This could be done via the use of photometric catalogs from real or simulated stellar surveys (\eg, \citealt{gaia2016}, \citealt{Robin2012}).
\item It is now well known that the contribution of clustering and environmental effects influence the colors (\eg, \citealt{madgwick2003}, \citealt{bamford09}) and spectral types \citep{zehavi2002} of galaxies: red and quiescent galaxies are mostly distributed in regions of high density, such as the centers of clusters, whereas blue and star forming galaxies are less clustered. Galaxy clustering also has an impact on source blending and confusion. These effects are not implemented in {\sc Stuff}, and this might bias our results in a way that is difficult to estimate. In order to limit this effect, one could select the areas of the analyzed survey that contain only field galaxies and use these areas as input data.
\item The present application uses as a reference the CFHTLS Deep survey, which sensitivity is very homogeneous over the field of view. This is, however, not the case for many surveys. A more general application of the method would require simulating each individual raw survey exposure, and performing the very same co-addition as with the observed data to generate  stacks, hence reproducing all the observational effects affecting the reduced images. However, this dramatically increases computing time and is currently out-of-reach except for the shallower surveys.
\end{itemize}

Because the pipeline in this work makes it possible to constrain not only the galaxy luminosity evolution, but also the evolution of galaxy sizes, it opens interesting perspectives for addressing the current debate on the evolution of galaxy sizes with cosmic time.
The contradictory results of, for example, \citet{Longhetti.etal:2007}, \citet{Trujillo.etal:2007}, \citet{Saracco10}, and \citet{Morishita2014} on the growth of massive early-type galaxies may be plagued with the varying selection effects in the surveys on which these analyses are based.

\section{Conclusions}
 
In the present article we lay the basis for a new method to infer robust constraints on galaxy evolution models. In this method, populations of synthetic galaxies are generated with the {\sc Stuff} empirical model, sampled from luminosity functions for each galaxy type, and determined by the SEDs of the bulge and disk components, and the B/T ratio. In order to reproduce the selection effects affecting real catalogs, we use the {\sc Skymaker} software to simulate realistic survey images with the appropriate instrumental  characteristics. Real and mock images undergo the same source extraction, using {\sc SExtractor}, and pre-processing pipeline. The distributions of extracted observables (fluxes and radii) are then compared, and we minimize their discrepancy using an adaptive MCMC sampling scheme in the parametric Bayesian indirect likelihood framework, designed for an efficient exploration of the parameter space. 

This is the first attempt in the field of galaxy evolution to make image simulation a central part of the inference process. We have tested the self-consistency of this approach using a simulated image of a CFHTLS Deep field covering 1 deg$^2$ on the sky in three bands: \textit{u} and \textit{i} in the optical, and $K_s$ in the near infrared, generated with the {\sc Stuff} model containing E/S0 and spiral galaxies with evolving size and luminosity. 

Starting from non-informative uniform priors, we find that our pipeline can reliably infer the input parameters governing the luminosity and size evolution of each of the galaxy populations in $\sim 10^4$ iterations, using few and disjointed observables, that is, the photometry (fluxes and radii) of the extracted sources in \textit{ui$K_s$}. In each test performed, the input parameters lie within the 68\% highest posterior density region.
 
We have also compared the results of our method with those of the classical photometric redshifts approach, with measurements from SED fitting on one of the mock sample, and found that when using the same set of observables (\textit{ui$K_s$} photometry), our inference pipeline yields more accurate results.
 
Now that the validity of our pipeline is established on mock data, we intend to apply it to the observed CFHTLS Deep fields. We could also combine these data with several extragalactic surveys at various depths and with different instrumental setups simultaneously, such as UDF \citep{Williams2010} at z $\sim$ 2, and SDSS \citep{Blanton2002} at z $\sim$ 0.1, in order to better constrain galaxies in a wide range of redshifts. Nevertheless, this application will raise various modeling issues. In particular, real survey images display a continuum of galaxy populations, and our model only generates a discrete number of galaxy populations, defined by their bulge and disk SEDs and their B/T ratio. In practice, the number of modeled populations will be limited by computing time, as more populations lead to more free parameters to infer, hence to more iterations for the pipeline to find the high probability regions. This will certainly require a compromise between the desired accuracy of the modeled universe and convergence of the chains within a reasonable computing time.
 
\begin{acknowledgements}
The authors thank Erwan Cameron for his useful comments on this work. Based on observations obtained with MegaPrime/MegaCam, a joint project of CFHT and CEA/IRFU, at the Canada-France-Hawaii Telescope (CFHT) which is operated by the National Research Council (NRC) of Canada, the Institut National des Science de l'Univers of the Centre National de la Recherche Scientifique (CNRS) of France, and the University of Hawaii. This work is based in part on data products produced at Terapix available at the Canadian Astronomy Data Centre as part of the Canada-France-Hawaii Telescope Legacy Survey, a collaborative project of NRC and CNRS. This work was partially supported by the ANR-13-BS05-002 SPIN(e) grant from the French Agence Nationale de la Recherche. 
\end{acknowledgements}
 
\begin{appendices}
\renewcommand\thetable{\thesection\arabic{table}}
\renewcommand\thefigure{\thesection\arabic{figure}}
 
\section{Derivation of the auxiliary likelihood function used in the present article}
\label{app:demolnL}
 
If one assumes that the number count in each bin $i$ is described by a Poisson distribution, the probability of $o_i$ given the model $s_i$ is:
\begin{align}
l_i= \frac{{e^{ - s_i } s_i ^{o_i} }}{{o_i!}}.
\end{align}
The likelihood function for the histogram is then:
\begin{align}
L = \prod_{i=1}^{b} \frac{{e^{ - s_i } s_i ^{o_i} }}{{o_i!}}.
\end{align}
 Correlations between adjacent bins are neglected here. The log-likelihood is therefore given by:
\begin{align}
\ln L = \sum_{i=1}^{b} (-s_i + o_i \ln(s_i) - \ln(o_i!)).
\end{align}
As we are interested in maximizing $\ln L$, $\ln(o_i!)$ is a constant that can be eliminated, so \textit{in fine}, we obtain \eq (\ref{lnL}). 
 
 
\section{Conversion LF parameters from \citet{faber07} to {\sc Stuff} parameters} \label{app:convfaberstuff}
 
In order to provide {\sc Stuff} with realistic LF parameters, we use \citet{faber07}, who used data from SDSS (\citealt{york2000}; \citealt{Blanton2002}), COMBO-17 (\citealt{wolf2001}; \citealt{wolf03}), 2dF \citep{Norberg2002}, and DEEP2 \citep{Davis2003} to derive the evolving LF parameters for two populations of red and blue galaxies. We associate the red and blue populations with our populations of E/S0 and spirals respectively. The LF parameters found by \citet{faber07} are listed in Table~\ref{tab:popsFaber07}, and apply to $z=0.5$.
\begin{table*}[ht]
\centering
\caption{\label{tab:popsFaber07} Luminosity function parameters of the blue and red populations of galaxies at $z=0.5$ inferred from SDSS, 2dF, COMBO-17, and DEEP2 data, adapted from Tables 3,4, and 6 of \citet{faber07}.}  
\begin{tabular}{c c c c c c c c c}
\hline
\hline
Population & $M^*_B$(z=0.5)& $\log_{10}(\phi^* [$Mpc$^{-3}])$(z=0.5) & P & Q & $\alpha$ \\
\hline
Red & -20.80 & -2.72 & -0.46 & -1.23 & -0.5 \\
Blue & -20.84 & -2.55 & 0.01 & -1.35 & -1.3 \\
\hline
\end{tabular}
\tablefoot{ The redshift evolution is fitted by $y=a_0(z=0.5)+a_1[\log_{10}(1+z)-\log_{10}(1+0.5)]/\log_{10}(2)$, where $M^*_B$ and $\log_{10}(\phi^*)$ are the zero points and Q and P are the slopes \resp $~$The LF parameters in \citealt{faber07} are given for $H_0$ = 70 km.$s^{-1}$.Mpc$^{-1}$}
\end{table*}
 In order to obtain the LF parameters for $z=0$, we use the fitted functions provided by \citet{faber07} for each population:
\begin{align}
M^*_B(z=0)=M^*_B(z)-\frac{Q \log_{10}(1+z)}{\log_{10}(2)}
\label{eq:Mbjohnson}
\end{align}                                   
\begin{align}
\log_{10}\phi^*(z=0)=\log_{10}\phi^*(z)-\frac{P \log_{10}(1+z)}{\log_{10}(2)}
\label{eq:Phiz05}
\end{align}  
The absolute magnitude in \eq \ref{eq:Mbjohnson} is given in the Johnson system. Because in our simulation {\sc Stuff} operates in the AB system, we use the AB offset calculated by \citet{frei94}:
\begin{align}
B_{AB}=B_{Johnson}-0.163
\end{align}
We then apply the transformation equations of \citet{jester05} for stars with $R_c-I_c < 1.15$ and $U-B > 0$, 
\begin{align}
B_{AB} = g + 0.39(g-r) + 0.21,
\end{align}
in order to derive g-band magnitudes: 
\begin{align}
M^*(z=0)_g=M^*_B(z=0)-0.39(g-r)-0.21-0.163.
\end{align}
We subsequently adopt average colors of $(g-r)_{E/S0}=0.75$ and $(g-r)_{Sp}=0.5$ from EFIGI data (de Lapparent, private communication) to derive the value of $M^*(z=0)_g$ for each population.
 
In {\sc Stuff}, the input LF parameters are provided assuming $H_0=100h~\mathrm{km}.\mathrm{s}^{-1}.\mathrm{Mpc}^{-3}$ with $h=1$. As \citet{faber07} provide their results assuming $h=0.7$, an additional conversion is needed:
\begin{align}
M^*_{STUFF} = M^*(z=0)_g - 5 \log_{10}h 
\end{align}
\begin{align}
\phi^*_{STUFF} = \phi^*h^{-3}.
\end{align}
In {\sc Stuff}, the LF evolution parameters are defined as:
 
\begin{align}
M^*(z)=M^*(z=0)+M_e \ln(1+z)
\label{eq:defmevol}
\end{align}
\begin{align}
\log_{10}\phi^*(z)=\log_{10}\phi^*(z=0)+\phi_e \log_{10}(1+z).
\label{eq:defphievol}
\end{align}
Combining  \eqs \ref{eq:Mbjohnson} and \ref{eq:Phiz05} with \eqs \ref{eq:defmevol} and \ref{eq:defphievol} respectively, we obtain:
 
\begin{align}
M_e=\frac{Q}{\ln(10) \log_{10}(2)}
\end{align}
\begin{align}
\phi_e=\frac{P}{\log_{10}(2)}.
\end{align}
\end{appendices}
The values of $P$ and $Q$ listed in Table~\ref{tab:popsFaber07} are used to derive the LF parameters for the populations of E/S0 and Sp. \textit{In fine}, the $\phi^*(z=0)$ of each population is reduced by a factor ten to limit computation time. The final LF parameters are listed in Table~\ref{tab:popparams} (\sct\ref{subsec:2_pop}).
 
\bibliographystyle{aa}
\bibliography{main_paper1}

\end{document}